\pdfoutput=1  

\documentclass[12pt]{article}

\usepackage{tocloft}
\usepackage{graphicx}
\usepackage{natbib}
\usepackage[protrusion=true,expansion=true]{microtype}
\usepackage{amsmath,amsthm,amsfonts,amssymb,mathrsfs,dsfont,mathtools}
\usepackage{booktabs}
\usepackage{url}
\usepackage{float}
\usepackage{color}
\usepackage{parskip}
\usepackage[explicit]{titlesec}
\usepackage{threeparttable}
\usepackage{setspace}
\usepackage[margin=0.85in]{geometry}
\usepackage[font = onehalfspacing]{caption}
\usepackage[usenames,dvipsnames]{xcolor}
\usepackage[hyperfootnotes=true]{hyperref}
\usepackage[labelfont=bf]{caption}
\usepackage{authblk}
\usepackage{caption}
\usepackage[titletoc,toc,title]{appendix}
\usepackage{enumerate}
\usepackage{soul}
\usepackage{subfig}
\usepackage{subfiles} 
\usepackage[capitalise, nameinlink]{cleveref}
\usepackage{dsfont}
\usepackage{chngpage} 
\usepackage{algorithm}
\usepackage{algpseudocode}

\newcommand\blfootnote[1]{%
	\begingroup
	\renewcommand\thefootnote{}\footnote{#1}%
	\addtocounter{footnote}{-1}%
	\endgroup
}

\newcommand{\E}{\mathbb{E}}

\usepackage{scalerel,stackengine}
\stackMath
\newcommand\reallywidehat[1]{%
	\savestack{\tmpbox}{\stretchto{%
			\scaleto{%
				\scalerel*[\widthof{\ensuremath{#1}}]{\kern-.6pt\bigwedge\kern-.6pt}%
				{\rule[-\textheight/2]{1ex}{\textheight}}
			}{\textheight}%
		}{0.5ex}}%
	\stackon[1pt]{#1}{\tmpbox}%
}

\usepackage{graphicx}
\newsavebox\CBox
\def\textBF#1{\sbox\CBox{#1}\resizebox{\wd\CBox}{\ht\CBox}{\textbf{#1}}}
\makeatletter
\newcommand*\bigcdot{\mathpalette\bigcdot@{.5}}
\newcommand*\bigcdot@[2]{\mathbin{\vcenter{\hbox{\scalebox{#2}{$\m@th#1\bullet$}}}}}
\makeatother

\setstcolor{red} 

\numberwithin{equation}{section}

\hypersetup{colorlinks = true, citecolor = MidnightBlue, linkcolor = MidnightBlue, urlcolor = MidnightBlue}

\titleformat{\section}{\normalfont\large\bfseries}{\thesection}{1em}{#1}
\titleformat{\subsection}{\normalfont\normalsize\bfseries}{\thesubsection}{1em}{#1}
\titleformat{\subsubsection}{\normalfont\normalsize\itshape}{\thesubsubsection}{1em}{#1}
\titlespacing\section{0pt}{12pt plus 4pt minus 2pt}{6pt plus 2pt minus 2pt}
\titlespacing\subsection{0pt}{12pt plus 4pt minus 2pt}{3pt plus 2pt minus 3pt}
\titlespacing\subsubsection{0pt}{12pt plus 4pt minus 2pt}{0pt plus 2pt minus 3pt}

\def\boxit#1{\vbox{\hrule\hbox{\vrule\kern6pt
			\vbox{\kern6pt#1\kern6pt}\kern6pt\vrule}\hrule}}

\definecolor{orange}{rgb}{1,0.5,0}
\definecolor{MyDarkBlue}{rgb}{0,0.08,0.45}

\newtheorem{remark}{Remark}[section]

\begin{document}
		\title{\Large \bfseries Deep equal risk pricing of financial derivatives with non-translation invariant risk measures
		\blfootnote{A GitHub repository with some samples of codes can be found at \href{https://github.com/alexandrecarbonneau}{github.com/alexandrecarbonneau}.}
		}
	
	\author[a] {Alexandre Carbonneau\thanks{Corresponding author.\vspace{0.2em} \newline
			{\it Email addresses:} \href{mailto:alexandre.carbonneau@mail.concordia.ca}{alexandre.carbonneau@mail.concordia.ca} (Alexandre Carbonneau), \href{mailto:frederic.godin@concordia.ca}{frederic.godin@concordia.ca} (Fr\'ed\'eric Godin).}}
	\author[b]{Fr\'ed\'eric Godin}
	\affil[a,b]{{\small Concordia University, Department of Mathematics and Statistics, Montr\'eal, Canada}}
	\vspace{-10pt}
	\date{ 
		\today}	
	\date{\bigskip\bigskip 
		\today}
	
	
	\maketitle \thispagestyle{empty} 
	
	
	\begin{abstract} 
		\vspace{-5pt}
		The use of non-translation invariant risk measures within the equal risk pricing (ERP) methodology for the valuation of financial derivatives is investigated. The ability to move beyond the class of convex risk measures considered in several prior studies provides more flexibility within the pricing scheme. In particular, suitable choices for the risk measure embedded in the ERP framework such as the semi-mean-square-error (SMSE) are shown herein to alleviate the price inflation phenomenon 
		observed under Tail Value-at-Risk based ERP as documented for instance in \cite{carbonneau2021equal}.		
		The numerical implementation of non-translation invariant ERP is performed through deep reinforcement learning, where a slight modification is applied to the conventional deep hedging training algorithm \citep[see][]{buehler2019deep} so as to enable obtaining a price through a single training run for the two neural networks associated with the respective long and short hedging strategies. The accuracy of the neural network training procedure is shown in simulation experiments not to be materially impacted by such modification of the training algorithm.
		
		\noindent \textbf{Keywords:} Finance, Option pricing, Hedging, Reinforcement learning, Deep learning.
		
	\end{abstract} 
		
	
	
	\doublespacing
	
	\setcounter{page}{1}
	\pagenumbering{arabic}

	\section{Introduction}
	\label{section:introduction}
	The equal risk pricing (ERP) methodology for derivatives valuation, which was initially proposed by \cite{guo2017equal}, entails setting the price of a contingent claim as the initial hedging portfolio value which leads to equal residual hedging risk for both the long and short positions under optimal hedges. This pricing procedure is associated with numerous advantageous properties, 
	such as the production of prices that are arbitrage-free under some technical conditions \citep[see][]{guo2017equal,marzban2020equal,carbonneau2021equal}, consistency with non-myopic global dynamic optimal hedging strategies, invariance of the price with respect to the position considered (i.e. long versus short), and the ability to consider general risk measures\footnote{For instance, the ability to depart from the quadratic penalty considered in the celebrated variance-optimal approach of \cite{schweizer1995variance} enables avoiding adverse behavior associated with the penalization of hedging gains.} for the objective function of the hedging optimization problem. 
	
	To further improve the ERP framework, several subsequent studies proposed some modifications to the original scheme. For instance, \cite{marzban2020equal} and \cite{carbonneau2021equal} use the physical probability measure rather than the risk-neutral one to perform hedging optimization; this has the advantage of improved interpretability of resulting prices on top of removing the subjectivity associated with the choice of the risk-neutral measure in an incomplete market setting. Furthermore,  to enhance the computational tractability of the ERP approach, these two studies also consider the set of convex risk measures to represent the risk exposure of hedged transaction for both long and short parties.\footnote{The original work from \cite{guo2017equal} considers expected penalties as risk measures, which do not possess all properties of convex risk measures (e.g. most lack the translation invariance property). For instance, the Tail-Value-at-Risk (TVaR) is not a particular case of an expected penalty.} Indeed, when convex measures are used, the translation invariance property leads to a useful characterization of equal risk prices which removes the need to perform a joint optimization over all possible hedging portfolio initial values.
	
	The most natural convex risk measure to consider within the ERP approach is arguably the Conditional Value-at-Risk (CVaR), which is equivalent to the Expected Shortfall (ES) or Tail-Value-at-Risk under the assumption that underlying loss variables are absolutely continuous. See \cite{rockafellar2002conditional} for a formal definition of the CVaR and a description of its properties. The $\text{CVaR}_\alpha$ can be interpreted as the operator computing a probability weighted average of worst-case risks occurring within an event of probability below or exactly $1-\alpha$, which is very intuitive. Moreover, it is a coherent risk measure in the sense of \cite{artzner1999coherent}, which implies favorable properties from a risk measurement standpoint.\footnote{The class of coherent risk measures is a subset of the class of convex risk measure which assumes for instance the subadditivity and positive homogeneity properties; the latter are more stringent than the convexity property satisfied by all convex risk measures.} Furthermore, the CVaR measure is used extensively in practice by the financial sector to quantify capital requirements, see for instance \cite{BCBS2016}.
	
	Due to its favorable properties, several studies use the CVaR within the ERP framework: see \cite{carbonneau2021equal} and \cite{carbonneau2021deep}. It was observed in the foremost that when only the underlying asset is used to hedge put options and conventional risk-neutral measures are used to determine the initial capital for hedging, the tail risk is much more pronounced for the short position than for the long one, especially for out-of-the-money puts. This leads to equal risk prices that are substantially higher than their risk-neutral counterparts when the confidence level $\alpha$ of the $\text{CVaR}_\alpha$ is high, to an extent that can cast doubt on the applicability of the method in practice. An avenue that was explored in the aforementioned study to remedy this drawback is to reduce the confidence level as prices were shown numerically to be positively related to the latter. Unfortunately, as shown in this present paper, reducing the confidence level to obtain smaller option prices becomes quickly impractical since the resulting hedging strategies exhibit poor risk mitigation performance with speculative behavior magnifying tail losses for very high quantiles above the CVaR confidence level.
	This approach should therefore not be pursued in practice. A second possible solution to the inflated ERP prices issue which is explored in \cite{carbonneau2021deep} consists in incorporating other hedging instruments (e.g. short-term options) within dynamic hedging schemes. That approach is shown therein to produce prices that are often still higher than the traditional risk-neutral ones, but much closer to them. This avenue was thus deemed successful when applicable. However, it requires a more sophisticated model to represent the price dynamics of hedging instruments, which complicates its implementation in practice. Furthermore, hedges relying on option trades might not be feasible or desirable under some circumstances (e.g. lack of liquidity).
	
	The aforementioned simulation-based results on ERP prices highlight the need to identify an ERP approach which can strictly rely on the underlying asset for hedging transactions and, at the same time, alleviate the price inflation obtained with CVaR-based ERP. A straightforward route to explore so as to attempt obtaining a satisfactory ERP method respecting the above constraints is to modify the risk measure acting as the objective function in the optimal hedging problems underlying the ERP framework. For instance, risk measures putting less relative weight on tail risk and more on more moderate risk scenarios should produce lower option prices. However, such risk measures (e.g. the semi-variance, semi-root-mean-square-error (SRMSE), etc.) do not necessarily satisfy properties of convex risk measures, in particular the translation invariance property. Equal risk prices stemming from such risk measure choices therefore do not have the convenient characterization associated with convex risk measures, which highlights the need of tailor-made numerical procedures handling this additional complexity.
	
	The main contribution of this manuscript is twofold. The first is to propose a modification of the deep reinforcement learning approach illustrated in \cite{carbonneau2021equal} and \cite{carbonneau2021deep} to handle non-translation invariant risk measures within ERP naturally and without excessive additional computational burden. This modification essentially consists in feeding varying initial hedging portfolio values with simulated risky asset paths to the deep hedging algorithm from \cite{buehler2019deep}, and then coupling the trained neural network output with a bisection search to seek the initial hedging portfolio value equating risks for both the long and short positions. The latter bisection method search has previously been suggested in a similar context for instance in \cite{marzban2020equal}. The training algorithm modification is shown in the present work not to lead to a material deterioration in the hedging performance of the neural network underlying the numerical approach.
	The second contribution consists in exploring equal risk prices of options generated when using typical non-translation invariant risk measures. It is seen that the use of the class of semi-$\mathbb{L}^{p}$ risk measures of the form $L(x) = x^p \mathds{1}_{ \{x>0\} }$ for $p > 0$ is able to reduce ERP prices to more natural levels better in line with these of existing methodologies while simultaneously resulting in effective trading policies. Indeed, numerical results indicate that equal risk prices generated by the class of semi-$\mathbb{L}^{p}$ risk measures can span wider ranges of prices than these obtained under the $\text{CVaR}_{\alpha}$ risk measures with conventional confidence level $\alpha$ values. The latter phenomenon is shown to hold across all moneyness levels for puts, and is robust to all risky asset dynamics considered. Furthermore, the benchmarking of neural networks trading policies hedging performance demonstrates that optimized policies under the semi-$\mathbb{L}^{p}$ objective functions are effective for mitigating hedging risk across all values of $p$ considered, where $p$ is shown to control the relative weight associated with extreme hedging losses. This is in contrast with the $\text{CVaR}_{\alpha}$ objective function where hedging policies optimized with relatively small confidence level $\alpha$ exhibit poor risk mitigation for loss quantiles larger than $\alpha$. Lastly, our results show that the use of the semi-$\mathbb{L}^{2}$ objective function to price long-term European puts with trades involving exclusively the underlying stock is almost as successful to reduce equal risk price values as compared to values obtained by trading shorter-term options with the CVaR$_{\alpha}$ risk measure. All of these results clearly demonstrate the benefit of using the class of semi-$\mathbb{L}^{p}$ risk measures within the ERP framework by simultaneously alleviating the price inflation phenomenon observed under the class of CVaR measures as well as resulting in effective trading policies for risk management.
	
	This paper is divided as follows. \cref{se:littrev} provides a literature review about incomplete market derivatives pricing, hedging methods and reinforcement learning in finance.
	The theoretical setting used for the ERP approach in the present work is presented in \cref{section:market_setup}. \cref{sec:methodology}  
	explains the reinforcement learning methodology for neural networks embedded in the ERP approach with the modified training algorithm proposed in this paper.
	\cref{section:numerical_results} displays results of numerical experiments associated with semi-$\mathbb{L}^{p}$ risk measures based ERP. \cref{section:conclusion} concludes.
		
	
		\section{Literature review}
		\label{se:littrev}
		
		Financial derivatives pricing in incomplete markets has received an extensive amount of attention in the literature. Numerous papers approach this problem through the selection of a suitable risk-neutral measure based on various considerations such as shifting of the drift to achieve risk-neutrality and model invariance, see \cite{hardy2001regime} and \cite{christoffersen2010option}, consistency with equilibrium models, see \cite{esw1994option} and \cite{duan1995garch}, or minimum entropy  distance between the physical and risk-neutral measures, see \cite{frittelli2000minimal}. Another strand of literature considers pricing methods consistent with optimal hedging strategies. At first, quadratic hedging methods were considered in \cite{follmer1988hedging}, \cite{schweizer1995variance}, \cite{elliott1998discrete} and \cite{bertsimas2001hedging} due to their tractability. However, as a consequence of the limitations associated to the quadratic penalty (e.g. penalizing equally gains and losses), other objective functions were considered in alternative dynamic hedging schemes such as quantile hedging \citep{follmer1999quantile}, expected penalty minimization \citep{follmer2000efficient} or VaR and CVaR optimization as in \cite{melnikov2012dynamic} and \cite{godin2016minimizing}. Some pricing schemes were also developed to enable consistency with non-quadratic hedging methods, for instance utility indifference \citep{hodges1989optimal} or risk indifference \citep{xu2006risk}. An issue with the latter approaches is that different prices are obtained depending on if a long or short position is considered in the derivative. The ERP approach developed by \cite{guo2017equal} identifying the derivative price equating hedged risk exposure of both long and short positions remedies this drawback by providing a unique price invariant to the direction (i.e. long versus short) of the position. Several additional papers have used or expanded on the initial ERP methodology. One problem often considered by that methodology is the tackling of market incompleteness arising from short-selling bans on the underlying asset: \cite{alfeus2019empirical}, \cite{ma2019pricing} and \cite{he2020revised}. \cite{marzban2020equal}  propose to substitute the risk-neutral measure for the physical measure during the determination of the equal risk price and to replace expected loss functions by convex risk measures within the objective function. \cite{carbonneau2021equal} provide a tractable methodology based on deep reinforcement learning to implement the ERP framework with convex risk measures under very general conditions.  
		\cite{carbonneau2021deep} examine the impact of introducing options as hedging instruments within the ERP framework under convex risk measures. 
		
		The computation of equal risk prices for derivatives is a highly non-trivial endeavor requiring advanced numerical schemes in most cases. \cite{marzban2020equal} propose to use dynamic programming which they apply on a robust optimization setting. Conversely, \cite{carbonneau2021equal} and \cite{carbonneau2021deep} use the deep reinforcement learning approach of \cite{buehler2019deep} coined as \textit{deep hedging}. Other papers have relied on the deep hedging methodology for the hedging of financial derivatives: \cite{cao2020discrete}, \cite{carbonneau2021deepIME}, \cite{horvath2021deep} and \cite{lutkebohmert2021robust}. Deep reinforcement learning is a very favorable technique for multistage optimization and decision-making in financial contexts: it allows tackling high-dimensional settings with multiple state variables, underlying asset dynamics and trading instruments. For this reason, it was used in multiple other works on derivatives pricing and hedging. Various techniques were considered such as Q-learning in  \cite{halperin2020qlbs} and \cite{cao2021deep}, proximal policy optimization in \cite{chong2021pseudo}, least squares policy iteration and fitted
		Q-iteration for American option pricing in \cite{li2009learning}, or batch policy gradient in \cite{buehler2019deep}. Moreover, various other financial problems were tackled through reinforcement learning procedures in the literature, for instance portfolio management as in \cite{moody1997optimization}, \cite{jiang2017deep}, \cite{pendharkar2018trading}, \cite{garcia2019continuous}, \cite{wang2020continuous}, \cite{ye2020reinforcement} and \cite{betancourt2021deep}, optimal liquidation, see \cite{bao2019multi}, or trading optimization as in \cite{hendricks2014reinforcement}, \cite{lu2017agent} and \cite{ning2018double}.
	

	\section{Financial market setup}
	\label{section:market_setup}
	
	This section details the mathematical framework for the financial market considered along with the theoretical setup for the ERP derivatives valuation approach.
	
	A discrete set of equally spaced time points spanning a horizon of $T$ years $\mathcal{T}\equiv\{0=t_0 < t_1 <\ldots<t_N=T\}$ with $t_n \equiv n \Delta$, $n=0,\ldots,N$ is considered. $\Delta$ corresponds to the length of a time period in years. Unless specified otherwise, the present study uses either $\Delta=1/260$ or $\Delta=1/12$ corresponding to daily or monthly periods.
	Moreover, consider the probability space $\left(\Omega, \mathcal{F}_N, \mathbb{P}\right)$ endowed with a filtration $\mathbb{F} \equiv \{ \mathcal{F}_n\}^N_{n=0}$ satisfying the usual conditions, with $\mathcal{F}_n$ being the sigma-algebra characterizing the information available to the investor at time $t_n$. Multiple traded assets are introduced in the financial market. First, a risk-free asset grows at a constant periodic risk-free rate $r \in \mathbb{R}$: its time-$t_n$ price is given by $B_{n} \equiv e^{r t_{n}}$.
	The $D+1$ other non-dividend paying risky asset prices are characterized by the vectorial stochastic processes $\{S^{(b)}_n\}^{N}_{n=0}$ and $\{S^{(e)}_n\}^{N-1}_{n=0}$ where $S^{(b)}_n \equiv \left[S^{(0,b)}_n, \ldots, S^{(D,b)}_n\right]$ and $S^{(e)}_n \equiv \left[S^{(0,e)}_n, \ldots, S^{(D,e)}_n\right]$ respectively represent the \textit{beginning-of-period} and \textit{end-of-period} prices of risky assets $0,\ldots,D$ available for trading at time $t_n$. This implies $S^{(b)}_n$ is $\mathcal{F}_n$-measurable (i.e. observable at time $t_n$) whereas $S^{(e)}_n$ is $\mathcal{F}_{n+1}$-measurable. Due to traded instruments changing on every time period (for example, some traded options mature contracts need to be rolled-over), it is possible to have $S^{(j,e)}_{n} \neq S^{(j,b)}_{n+1}$, $j=1,\ldots,D$. However, the risky asset $j=0$ is assumed to be an underlying asset with no maturity such as a stock, thus available for trading on all periods. Hence, $S^{(0,e)}_{n} = S^{(0,b)}_{n+1}$. For simplicity, an absence of market frictions is assumed throughout the paper. Correspondingly, it is assumed all positions in a given portfolio are liquidated at the end of any period, and are repurchased at the beginning of the next if needed.
	
	A European-type derivative of time-$t_N$ payoff $\Phi\left(S^{(0,b)}_{N}\right)$ is considered. A suitable price for that contract and corresponding hedging strategies must be determined. Define a trading strategy $\delta \equiv \{ \delta_n \}^N_{n=0}$ as an $\mathbb{F}$-predictable process\footnote{This means $\delta_0$ is $\mathcal{F}_0$-measurable and $\delta_n$ is $\mathcal{F}_{n-1}$-measurable for $n=1,\ldots,N$.} where $\delta_n \equiv \left[\delta^{(0)}_{n}, \ldots, \delta^{(D)}_{n}, \delta^{(B)}_{n}\right]$. The latter comprises $\delta^{(0:D)}_{n} \equiv \left[\delta^{(0)}_{n}, \ldots, \delta^{(D)}_{n}\right]$ which contains the positions in all respective risky assets $0,\ldots,D$ within the portfolio between time $t_{n-1}$ and time $t_n$, and $\delta^{(B)}_{n}$ which contains the portfolio investment in the risk-free asset for the same period. For a trading strategy $\delta$, the corresponding time-$t_n$ portfolio value is defined as
	\begin{equation*}
	V^\delta_n \equiv  \begin{cases}
	\delta^{(0:D)}_0 \bigcdot S^{(b)}_{0} + \delta^{(B)}_0 B_0, \quad n=0,
	\\ \delta^{(0:D)}_n \bigcdot S^{(e)}_{n-1} + \delta^{(B)}_n B_n, \quad n=1,\ldots,N,
	\end{cases}
	\end{equation*}
	where $\bigcdot$ is the conventional dot product. 	
	A trading strategy $\delta$ is said to be \textit{self-financing} if 
	\begin{equation*}
	\delta^{(0:D)}_{n+1} \bigcdot S^{(b)}_{n} + \delta^{(B)}_{n+1} B_n = V^\delta_n, \quad n=0,\ldots,N-1.
	\end{equation*}
	Denote by $\Pi$ the set of all self-financing trading strategies that are sufficiently well-behaved mathematically.\footnote{Details characterizing well-behavedness in the context of the present study are omitted to avoid lengthy discussions straying us away from the main research objectives of the present work.} It turns out that the portfolio value process of self-financing trading strategies can be expressed conveniently in terms of so-called \textit{discounted gains}. For a trading strategy $\delta \in \Pi$, the latter are defined as
	\begin{equation*}
	G^\delta_0 \equiv 0, \quad G^\delta_n \equiv \sum_{j=1}^{n} \delta^{(0:D)}_j \bigcdot \left(   B^{-1}_j S^{(e)}_{j-1} - B^{-1}_{j-1} S^{(b)}_{j-1}\right), \quad n=1,\ldots,N.
	\end{equation*}
	Using standard arguments outlined for instance in \cite{lamberton2007introduction}, for any self-financing trading strategy $\delta \in \Pi$,
	\begin{equation*}
	V^\delta_n = B_n \left(V^\delta_0 + G^\delta_n\right).
	\end{equation*}
	Such representation is convenient as it allows avoiding calculating $\delta^{(B)}_n$ for $n=0,\ldots,N$ explicitly when calculating the portfolio value.
	
	Aforementioned definitions allow posing the main optimization problems underlying the ERP methodology, which consist in finding the best self-financing trading strategies leading to optimal hedges in terms of penalized hedging errors at the maturity of the derivative. Solutions of such problems are referred to as \textit{global hedging procedures} due to their measurement of hedging efficiency in terms of risk at maturity rather than on a period-by-period basis. Consider a given risk measure $\rho$ characterizing the risk aversion of the hedger.\footnote{A risk measure is a mapping taking a random variable representing a random loss as input, and return a real number representing its perceived risk as an output.} Specific examples of risk measures considered in this study are formally defined subsequently. For a given value of $V_0 \in \mathbb{R}$, define mappings $\epsilon^{(\mathcal{L})}:\mathbb{R} \rightarrow \mathbb{R}$ and $\epsilon^{(\mathcal{S})}:\mathbb{R} \rightarrow \mathbb{R}$ representing optimal residual hedging risk respectively for a long or short position in the derivative when the initial portfolio value is $V^\delta_0 = V_0$ as	
	\begin{align}
	\epsilon^{(\mathcal{L})}(V_0) &\equiv \underset{\delta\in \Pi}{\min} \, \rho \left(-\Phi(S^{(0,b)}_{N}) -V^\delta_{N}\right), \quad	 \epsilon^{(\mathcal{S})}(V_0) \equiv \underset{\delta\in \Pi}{\min} \, \rho \left(\Phi(S^{(0,b)}_{N}) -V^\delta_{N}\right). \label{eq:risk_longshort}
	\end{align}
	Optimal hedging strategies are the minimizing arguments of such optimization problems:
	\begin{align*}
		\delta^{(\mathcal{L})}(V_0) &\equiv \underset{\delta\in \Pi}{\arg\min} \, \rho \left(-\Phi(S^{(0,b)}_{N}) -V^\delta_{N}\right), \quad	 \delta^{(\mathcal{S})}(V_0) \equiv \underset{\delta\in \Pi}{\arg\min} \, \rho \left(\Phi(S^{(0,b)}_{N}) -V^\delta_{N}\right). 
		\end{align*}
	This leads to the definition of the \textit{equal risk price} $C^*_0$ of the derivative $\Phi$ as the initial portfolio value $V_0$ such that the optimal residual hedging risk is equal for both the long and short positions, i.e.
	\begin{equation}
	\label{ERPdef}
	\epsilon^{(\mathcal{L})}(-C^*_0) = \epsilon^{(\mathcal{S})}(C^*_0). 
	\end{equation}
	Conditions on $\rho$ have to be imposed to guarantee the existence and uniqueness of the equal risk price (e.g. monotonicity of $\rho$). Under the assumption that $\rho$ is a convex risk measure, \cite{carbonneau2021equal} provide sufficient conditions to obtain existence and uniqueness of the solution to \eqref{ERPdef}, see Theorem $2.1$ of the latter paper.
	
	
	\begin{remark}
	\label{remark:convex_ERP}
	Under a convex measure $\rho$, \cite{marzban2020equal} and \cite{carbonneau2021equal} also obtain the following characterization of the equal risk price
	\begin{align}
	C^*_0 = 0.5B_N \left(\epsilon^{(\mathcal{S})}(0)-\epsilon^{(\mathcal{L})}(0)\right). \label{eq:ref_ERP_convex}
	\end{align}
	 Representation \eqref{eq:ref_ERP_convex} is very convenient as it requires to only obtain the optimal residual risk exposure when the initial portfolio is null instead of having to iteratively try multiple initial portfolio values. However, when $\rho$ is not translation invariant, such representation does not hold anymore, and a tailor-made numerical scheme must thus be developed to solve for the root-finding problem \eqref{ERPdef}.
	\end{remark}
	
	The present work aims among others at examining a class of non-translation invariant risk measures.
	The main class of risk measures under study will be referred to as the \textit{semi-$\mathbb{L}^p$ risk measures}, which are defined as
	\begin{align}
	\rho(X) \equiv \mathbb{E}\left[ X^p \mathds{1}_{ \{X>0\}}\right]^{1/p}, \quad p > 0. \label{eq:ref_Lp_risk_measure}
	\end{align}
	The latter risk measure is clearly monotonous (i.e. $X \geq Y$ almost surely implies $\rho(X) \geq \rho(Y)$), but lacks the translation invariance property. One important advantageous property of this class of risk measures is in penalizing exclusively hedging losses, not gains. Furthermore, the parameter $p$ acts as a risk aversion barometer as higher values of $p$ put more relative weight on higher losses.
	
	The CVaR measure is also considered in some experiments of the present paper for benchmarking purposes as it is used in \cite{carbonneau2021equal} and \cite{carbonneau2021deep}. Such a risk measure can be formally defined as
	\begin{eqnarray*}
	\text{VaR}_\alpha (X) \equiv \inf \{ x : \mathbb{P}[X \leq x] \geq \alpha \}, \quad \text{CVaR}_\alpha (X) \equiv \frac{1}{1-\alpha}\int_{\alpha}^{1} \text{VaR}_\gamma (X) d \gamma
	\end{eqnarray*}
	for a confidence level $\alpha$ in $(0,1)$. Whenever $X$ is an absolutely continuous random variable, the CVaR admits the intuitive representation $\text{CVaR}_\alpha (X) = \mathbb{E}\left[X | X \geq \text{VaR}_\alpha (X)\right]$. The CVaR is a coherent risk measure as shown \cite{rockafellar2002conditional}, which implies it satisfies the monotonicity and translation invariance properties. 
	
	
	
	\section{Methodology}
	\label{sec:methodology}
	
	The present section details the reinforcement learning approach followed to solve the optimization problems underlying the ERP methodology. The approach consists in applying the deep hedging algorithm of \cite{buehler2019deep} by representing hedging policies with neural networks. A slight modification to the latter paper's training methodology is required to solve the ERP global hedging problems when the risk measure is not translation invariant. An accuracy assessment is performed for the modified training algorithm.
	
	\subsection{Neural network approximation of the optimal solution}
	The approach followed to obtain a numerical solution to the optimization problems \eqref{eq:risk_longshort} is based on a parametric approximation of the trading policy with a neural network trained using reinforcement learning. The general idea is as follows. In multiple setups, especially those involving Markovian dynamics, the optimal trading strategies $\delta^{(\mathcal{S})}(V_0)$ and $\delta^{(\mathcal{L})}(V_0)$ often admit the following functional representation for some functions $\tilde{\delta}^{(\mathcal{L})}$ and $\tilde{\delta}^{(\mathcal{S})}$:
	\begin{equation}
	\label{eq:optPolFunc}
	\delta^{(\mathcal{L})}_{n+1}(V_0) = \tilde{\delta}^{(\mathcal{L})} \left(T - t_n, S_n^{(b)},V_n,\mathcal{I}_n\right), \quad \delta^{(\mathcal{S})}_{n+1}(V_0) = \tilde{\delta}^{(\mathcal{S})} \left(T - t_n, S_n^{(b)},V_n,\mathcal{I}_n\right), \quad n=0,\ldots,N-1,
	\end{equation}
	 where $\delta^{(\mathcal{L})}_{n+1}(V_0)$ and $\delta^{(\mathcal{S})}_{n+1}(V_0)$ are to be understood as the optimal time-$t_n$ hedges for the long and short position when time-$0$ capital investment is $V_0$, and $\mathcal{I}_n$ is a $\mathcal{F}_n$-measurable random vector containing a set of additional state variables summarizing all necessary information to make the optimal portfolio rebalancing decision. For instance, $\mathcal{I}_n$ can contain underlying asset volatilities if the latter asset has a GARCH dynamics \citep[see][]{augustyniak2017assessing}, current probabilities of being in the various respective regimes when in a regime-switching setup \citep[see][]{franccois2014optimal}, implied volatilities when options are used as hedging instruments \citep[see][]{carbonneau2021deep}, current assets positions when in the presence of transaction costs \citep[see][]{breton2017global}, and so on. 
	 
	 The functional representation \eqref{eq:optPolFunc} enables the approximation of the optimal policies as parameterized functions. The class of functions considered in this paper is the classical \textit{feedforward neural network} (FFNN) class, which is formally defined subsequently. Indeed, two distinct FFNNs are used to approximate the optimal trading policy of the long and short parties by mapping inputs $\{T - t_n,S_n^{(b)},V_n,\mathcal{I}_n\}$ into the respective (long or short) portfolio positions of risky assets $\delta_{n+1}^{(0:D)}$ for any $n=0,\ldots,N-1$.\footnote{Recall that since the trading strategy is self-financing, $\delta^{(B)}_{n+1}$ is characterized by $\delta^{(0:D)}_{n+1}$ and $V_n$.}
	More precisely, denote by $F_{\theta}^{(\mathcal{L})}$ and $F_{\theta}^{(\mathcal{S})}$ the neural network mappings for respectively the long and short trading positions where $\theta \in \mathbb{R}^q$ is the $q$-dimensional set of parameters of the FFNNs.\footnote{
	While the neural network architecture of $F_{\theta}^{(\mathcal{L})}$ and  $F_{\theta}^{(\mathcal{S})}$ considered in this paper is the same for both neural networks in terms of the number of hidden layers and neurons per hidden layer, and thus the total number $q$ of parameters to fit is the same for both neural networks, one could also consider two different architectures for $F_{\theta}^{(\mathcal{L})}$ and  $F_{\theta}^{(\mathcal{S})}$ with no additional difficulty.
	} For a given parameter set $\theta$ distinct for each neural network, the associated trading strategies are given by 
	\begin{equation*}
	\delta^{(\mathcal{L}, \theta)}_{n+1}(V_0) \equiv F_{\theta}^{(\mathcal{L})} \left(T-t_n,S_n^{(b)},V_n,\mathcal{I}_n\right), \quad
	\delta^{(\mathcal{S}, \theta)}_{n+1}(V_0) \equiv F_{\theta}^{(\mathcal{S})} \left(T-t_n,S_n^{(b)},V_n,\mathcal{I}_n\right), \quad n=0,\ldots,N-1.
	\end{equation*}
	The optimization of trading strategy in problem \eqref{eq:risk_longshort} is thus replaced by the optimization of neural network parameters $\theta$ according to
	\begin{align}
	\tilde{\epsilon}^{(\mathcal{L})}(V_0) \equiv \underset{\theta \in \mathbb{R}^{q}}{\min} \, \rho \left(-\Phi(S^{(0,b)}_{N}) - V_N^{\delta^{(\mathcal{L}, \theta)}} \right), \quad \tilde{\epsilon}^{(\mathcal{S})}(V_0) \equiv \underset{\theta \in \mathbb{R}^{q}}{\min} \, \rho \left(\Phi(S^{(0,b)}_{N}) -V^{\delta^{(\mathcal{S}, \theta)}}_N \right). \label{eq:risk_long_short_NNet}
	\end{align}
	Note that the set of optimal parameters $\theta$ will be different for the long and the short trading strategies. Furthermore, problems \eqref{eq:risk_long_short_NNet} only lead to an approximate solution to the initial problems \eqref{eq:risk_longshort} since the FFNNs are approximations of the true functional representation $\tilde{\delta}^{\mathcal{(L)}}$ and $\tilde{\delta}^{\mathcal{(S)}}$.
	Nevertheless, by relying on the universal approximation property of FFNNs \citep[see for instance][]{hornik1991approximation}, \cite{buehler2019deep} show that there exist neural networks such that the solution $\tilde{\epsilon}^{(\mathcal{L})},\tilde{\epsilon}^{(\mathcal{S})}$ from \eqref{eq:risk_long_short_NNet} can be made arbitrarily close to the solution $\epsilon^{(\mathcal{L})},\epsilon^{(\mathcal{S})}$ from \eqref{eq:risk_longshort}. 
	
	The mathematical definition of FFNNs architecture is now provided.
	For $L, d_0, \ldots, d_{L+1} \in \mathbb{N}$, let $F_{\theta}:\mathbb{R}^{d_0} \rightarrow \mathbb{R}^{d_{L+1}}$ be a FFNN:
	\begin{align}
	F_{\theta}(X)&\equiv o \circ h_{L} \circ \ldots \circ h_{1}, \nonumber
	\\ \quad h_l(X) &\equiv g(W_l X + b_l), \quad l = 1, \ldots, L, \nonumber
	\\ \quad o(X) &\equiv W_{L+1}X + b_{L+1}, \nonumber
	\end{align}
	where $\circ$ denotes the function composition operator. Thus, $F_\theta$ is a composite function of $h_1, \ldots, h_L$ commonly known as \textit{hidden layers} which each apply successively an affine and a nonlinear transformation to input vectors, and also of the \textit{output function} $o$ applying an affine transformation to the last hidden layer. The set of parameters $\theta$ to be optimized consists of all weight matrices $W_l \in \mathbb{R}^{d_{l} \times d_{l-1}}$ and bias vectors $b_l \in \mathbb{R}^{d_l}$ for $l=1,\ldots,L+1$.
		
	\subsection{Calibration of neural networks through reinforcement learning}
	\label{subsec:calibration_NNETS}
	As in \cite{buehler2019deep}, the training of neural networks in this paper relies on a stochastic policy gradient algorithm, also known as actor-based reinforcement learning. This class of procedures optimizes directly the policy (i.e. the actor) parameterized as a neural network with minibatch stochastic gradient descent (SGD) so as to minimize a cost function as in \eqref{eq:risk_long_short_NNet}. Without loss of generality, the training algorithm is hereby only provided for the neural network $F_{\theta}^{(\mathcal{S})}$ associated with the short position, as steps for the long position are entirely analogous. 
	
	\subsubsection{Fixed and given $V_0$ case}
	\label{subsubsec:fixed_V0}
	The training procedure to calibrate $\theta$ is first described for a fixed and given initial capital investment $V_0$ as originally considered in \cite{buehler2019deep}. A slight modification to the algorithm will subsequently be presented in \cref{subsubsec_nontrans_inv} to tackle the non-translation invariant risk measure case studied in this paper.
	Let $J : \mathbb{R}^{q} \times \mathbb{R} \rightarrow \mathbb{R}$ be the cost function for the short position hedge:
	\begin{align}
	J(\theta, V_0) \equiv \rho \left(\Phi(S^{(0,b)}_{N}) -V^{\delta^{(\mathcal{S}, \theta)}}_{N}\right), \quad \theta \in \mathbb{R}^{q}, V_0 \in \mathbb{R}. \label{eq:ref_cost_func}
	\end{align}
	
	The parameters set $\theta$ is sequentially refined to produce a sequence of estimates $\{\theta_j\}_{j\geq1}$ minimizing the cost function $J$ over time. This iterative procedure is as follows. First, parameters of the neural network are initialized with the Glorot
	uniform initialization of \cite{glorot2010understanding}, which gives the initial value of the sequence $\theta_0$. 
	Then, to start refining the parameters, a set of $M=400,\!000$ paths containing traded asset values and other exogenous variables associated with the assets dynamics is generated by Monte Carlo simulation. The set of such paths is referred to as a \textit{training set}. On each iteration of SGD, i.e. on each update of $\theta_j$ to $\theta_{j+1}$,  a minibatch consisting in a subset of size $N_{\text{batch}}=1,\!000$ of paths from the training set is used to estimate the cost function in \eqref{eq:ref_cost_func}. More precisely, for $\theta = \theta_j$, $F_\theta^{(\mathcal{S})}$ is used to compute the assets positions at each rebalancing date and for each path within the minibatch. Let $\mathbb{B}_j \equiv \{\pi_{i,j}\}_{i=1}^{N_{\text{batch}}}$ be the resulting set of hedging errors from this minibatch, where $\pi_{i,j}$ is the $i$th hedging error when $\theta = \theta_j$. Then, for $\hat{\rho} :\mathbb{R}^{N_{\text{batch}}} \rightarrow \mathbb{R}$ the empirical estimator of $\rho(\pi)$ evaluated with $\mathbb{B}_j$, the update rule for $\theta_j$ to $\theta_{j+1}$ is
	$$\theta_{j+1} = \theta_{j} - \eta_j \nabla_\theta \widehat{\rho}(\mathbb{B}_j),$$
	where $\{\eta_j\}_{j \geq 1}$ are small positive real values and $\nabla_\theta$ denotes the gradient operator with respect to $\theta$. For instance, under the semi-$\mathbb{L}^p$ class of risk measures which is extensively studied in the numerical section, the empirical estimator has the representation
	\begin{align}
	\widehat{\rho}\left(\mathbb{B}_j\right) \equiv \left(\frac{1}{N_{\text{batch}}} \sum_{i=1}^{N_{\text{batch}}} \pi_{i,j}^p \mathds{1}_{ \{\pi_{i,j}>0\} }\right)^{1/p}. \nonumber
	\end{align}
	Lastly, the computation of the gradient of the empirical cost function with respect to $\theta$ can be done explicitly with modern deep learning libraries such as Tensorflow \citep{abadi2016tensorflow}. Also, the Adam optimizer \citep{kingma2014adam} can be used to dynamically determined the $\eta_j$ values. The following section presents the modification to the training algorithm proposed in this paper to compute equal risk prices under non-translation invariant risk measures.
	
	\subsubsection{Non-translation invariant risk measures case}
	\label{subsubsec_nontrans_inv}
	The main objective of this paper is to study the valuation of financial derivatives with the ERP framework under non-translation invariant risk measures. This requires solving the root-finding problem of the initial portfolio value $V_0$ that equates $\tilde{\epsilon}^{(\mathcal{L})}(-V_0)$ and $\tilde{\epsilon}^{(\mathcal{S})}(V_0)$; this study considers a bisection scheme for such a purpose. However, one important drawback of the bisection algorithm in the context of this paper is the requirement to obtain multiple evaluations of $\tilde{\epsilon}^{(\mathcal{L})}(-V_0)$ and $\tilde{\epsilon}^{(\mathcal{S})}(V_0)$ for different values of $V_0$, which can be very costly from a computational standpoint. One naive approach to implement the bisection algorithm is to proceed as follows:
	\begin{itemize}
		\item [1)] For a given value of $V_0$, train the long and short neural networks $F_{\theta}^{(\mathcal{S})}$ and $F_{\theta}^{(\mathcal{L})}$ on the training set.
		\item [2)] Evaluate the optimal residual hedging risk $\tilde{\epsilon}^{(\mathcal{S})}(V_0)$ and $\tilde{\epsilon}^{(\mathcal{L})}(-V_0)$ with $F_{\theta}^{(\mathcal{S})}$ and $F_{\theta}^{(\mathcal{L})}$ on a \textit{test set} of $100,\!000$ additional independent simulated paths.
		\item [3)] If $\Delta(V_0) \equiv \tilde{\epsilon}^{(\mathcal{S})}(V_0) - \tilde{\epsilon}^{(\mathcal{L})}(-V_0) \approx 0$ according to some closeness criterion, then $C_0^{\star} = V_0$ is the equal risk price. Otherwise, update $V_0$ with the bisection algorithm and go back to step $1)$.
	\end{itemize}
	The important drawback of this naive approach lies in the necessity to retrain $F_{\theta}^{(\mathcal{S})}$ and $F_{\theta}^{(\mathcal{L})}$ for each iteration of the bisection algorithm in step $1$. To circumvent the latter pitfall, this study proposes to slightly modify the training algorithm such that the neural networks learn the optimal mappings not only for \textit{a unique fixed} initial capital investment, but rather for an \textit{interval} of values for $V_0$. This provides the important benefit of only having to train $F_{\theta}^{(\mathcal{S})}$ and $F_{\theta}^{(\mathcal{L})}$ once, which thus circumvents the previously described computational burden. 
	
	The slight modification made to the training algorithm described in \cref{subsubsec:fixed_V0} is now described. At the beginning of each SGD step, on top of sampling a minibatch of paths of risky assets, the value of $V_0$ is also randomly sampled within the initial interval of values used for the bisection algorithm. For instance, in numerical experiments conducted in \cref{section:numerical_results}, the initial interval considered for the bisection algorithm is $[0.75C_0^{\mathbb{Q}}, 1.50C_0^{\mathbb{Q}}]$ where $C_0^{\mathbb{Q}}$ is the risk-neutral price of $\Phi$ under a chosen conventional equivalent martingale measure $\mathbb{Q}$.\footnote{
	If the equal risk price is outside the initial search interval $[0.75C_0^{\mathbb{Q}}, 1.50C_0^{\mathbb{Q}}]$, the bisection algorithm must be applied once again with a new initial search interval, and the neural networks $F_{\theta}^{(\mathcal{S})}$ and $F_{\theta}^{(\mathcal{L})}$ must be trained once again on this new interval.
	} 
	This approach is simple to implement as it naturally leverages the fact that portfolio values are already used within input vectors of the neural networks. However, it should be noted that learning the optimal hedge for various initial capital investments is more complex, and thus a more challenging task for neural networks as compared to learning the optimal trading policy for a fixed $V_0$. Nevertheless, Monte Carlo experiments provided in \cref{appendix:proof_convergence_results} show that incorporating this slight modification to the training algorithm does not materially impact the optimized neural networks performance. 

	Pseudo-codes of the training and bisection procedures are presented respectively in \cref{algo:pseudo_code_training} and \cref{algo:pseudo_code_bisection} of \cref{appendix:pseudo_code}. An implementation in Python and Tensorflow to replicate numerical experiments presented in \cref{section:numerical_results} can also be found online at \href{https://github.com/alexandrecarbonneau}{github.com/alexandrecarbonneau}.
	
	\begin{remark}
		In numerical experiments of \cref{section:numerical_results}, the benchmarking of equal risk prices generated under the class of semi-$\mathbb{L}^{p}$ risk measures to the ones obtained with a class of convex risk measures, namely the CVaR, is performed. The numerical scheme used to obtain equal risk prices under the CVaR$_{\alpha}$ risk measure follows the methodology of \cite{carbonneau2021equal} by evaluating $C_0^{\star}$ with \eqref{eq:ref_ERP_convex} where $\tilde{\epsilon}^{(\mathcal{L})}(0)$ and $\tilde{\epsilon}^{(\mathcal{S})}(0)$ are computed with the steps of \cref{subsubsec:fixed_V0} with $V_0 = 0$ and with the empirical estimator of $\rho(\pi)$ as
		$$\widehat{\rho}(\mathbb{B}_j) = \widehat{\text{VaR}}_{\alpha}(\mathbb{B}_{j}) + \frac{1}{(1-\alpha)N_{\text{batch}}}\sum_{i=1}^{N_{\text{batch}}}\max(\pi_{i,j}-\widehat{\text{VaR}}_{\alpha}(\mathbb{B}_{j}),0),$$
		where $\widehat{\text{VaR}}_{\alpha}(\mathbb{B}_{j})$ is the usual empirical estimator of the Value-at-Risk statistic with the sample $\mathbb{B}_{j}$ at level $\alpha$.
		
	\end{remark}

	\begin{remark}
		For all numerical experiments under the semi-$\mathbb{L}^{p}$ risk measure conducted in this paper, a preprocessing of the feature vectors is applied, using $\{T-t_n, \log(S_n^{(b)}/K), V_n/\tilde{V}, \mathcal{I}_n\}$ instead of $\{T - t_n, S_n^{(b)},V_n,\mathcal{I}_n\}$ where $\tilde{V}$ is defined as the midpoint value of the initial search interval of the bisection algorithm $[V_A, V_B]$, i.e. $\tilde{V} \equiv 0.5(V_A + V_B)$. Note that \cite{carbonneau2021equal} and \cite{carbonneau2021deep} consider similar preprocessing for risky asset prices, while \cite{carbonneau2021deepIME} considers a similar preprocessing for portfolio values. Furthermore, under the CVaR$_{\alpha}$ objective function, the same preprocessing for risky asset prices is used, but portfolio values are not preprocessed as the bisection algorithm is not required to be used in this case, i.e. $V_n$ rather than $V_n/\tilde{V}$ is used in feature vectors.
	\end{remark}

	Lastly, it is worth highlighting an additional advantage from a computational standpoint of the class of semi-$\mathbb{L}^{p}$ objective functions described in this paper over the CVaR$_{\alpha}$ measures as considered for instance in \cite{carbonneau2021equal} and \cite{carbonneau2021deep} when relying on the neural network-based hedging scheme. Indeed, under the CVaR$_{\alpha}$ objective function, the use of minibatch stochastic gradient descent procedures to train neural networks restrain the use of extremely large quantiles for the CVaR$_{\alpha}$ (for instance, larger values than $0.99$). The latter stems from the following observations. From a statistical standpoint, the estimation variance of CVaR$_{\alpha}$ increases with $\alpha$. Furthermore, the empirical estimator of CVaR$_{\alpha}$ is biased in finite sample size, whereas the empirical estimator of the semi-$\mathbb{L}^{p}$ risk measure is unbiased for any sample size. However, while larger minibatches would provide a more accurate estimate of the gradient, i.e. reduce the variance and the bias of the CVaR estimator, this is not necessarily a favorable avenue for training neural networks. Indeed, as noted in \cite{goodfellow2016deep}, the amount of memory required by hardware setups can be a limiting factor to increasing minibatch size. Furthermore, most SGD algorithms converge faster in terms of total computation when allowed to approximate gradients faster (i.e. with smaller samples and more SGD steps). The interested reader is referred to Chapter $8.1.3$ of \cite{goodfellow2016deep} for additional information about the implications of the minibatch size on SGD procedures. This computational pitfall of pairing stochastic gradient descent with extreme values of $\alpha$ under the CVaR$_{\alpha}$ measure is not present under the semi-$\mathbb{L}^{p}$, which further motivates its use in the context of equal risk pricing and optimal hedging.
	
	\section{Numerical experiments}
	\label{section:numerical_results}
	This section presents several numerical experiments conducted to investigate prices produced by the ERP methodology under different setups. The common theme of all experiments is to examine option prices generated by the ERP framework under the class of semi-$\mathbb{L}^{p}$ risk measures. 
	The analysis starts in \cref{subsec:sens_analysis} with a sensitivity analysis of equal risk prices with respect to the choice of objective function. This is carried out by comparing $C_0^{\star}$ generated with the CVaR$_{\alpha}$ and semi-$\mathbb{L}^{p}$ across different values of $\alpha$ and $p$ controlling the risk aversion of the hedger. The hedging performance of embedded neural networks hedging policies obtained under these objective functions is also assessed. Moreover, a sensitivity analysis with respect to the choice of underlying asset price dynamics is carried out in \cref{subsec:diff_dyn} so as to test the impact of the inclusion of jump or volatility risk. Lastly, \cref{subsec:option_hedges} presents the benchmarking of equal risk prices for long maturity options obtained under the semi-$\mathbb{L}^{p}$ risk measures with trades involving exclusively the underlying stock against these generated with option hedges under the CVaR$_{\alpha}$ objective function.
	
	\subsection{Experiments setup}
	\label{subsec:market_setup}
	Unless specified otherwise, the option to price and hedge is a European put with payoff $\Phi(S_N^{(0,b)}) \equiv \max(K - S_N^{(0,b)}, 0)$ of maturity of $T=60/260$ and strike price $K$. Daily hedges with the underlying stock are used (i.e. $N=60$). The use of option hedges and different maturities for $\Phi$ is considered exclusively is \cref{subsec:option_hedges}. Furthermore, the stock has an initial price of $S_0^{(0,b)} = 100$ and the annualized continuous risk-free rate is set at $r = 0.02$. Different moneyness levels are considered with $K=90,100$ and $110$ for respectively out-of-the-money (OTM), at-the-money (ATM), and in-the-money (ITM) puts. 
	
	Moreover, as described in \cref{sec:methodology}, two distinct feedforward neural networks are considered for the functional representation of the long and short hedging policies. The architecture of every neural networks is a FFNN of two hidden layers ($L=2$) with $56$ neurons per layer ($d_1 = d_2 = 56$). The activation function considered is the well-known rectified linear activation function (ReLU) with $g(x) \equiv \max(x,0)$. For the training procedure, a training set of $400,\!000$ paths is simulated with the $\mathbb{P}$-dynamics of the underlying stock. A total of $100$ epochs\footnote{
		An epoch is defined as a complete iteration of the training set with stochastic gradient descent. For example, for a training set of $400,\!000$ paths and a minibatch size of $1,\!000$, one epoch consists of $400$ updates of the set of trainable parameters $\theta$. 
	} is used with a minibatch size of $1,\!000$ sampled exclusively from the training set. The Adam optimizer with a learning rate hyperparameter of $0.0005$ is used with Tensorflow for the implementation of the stochastic gradient descent procedure. Also, all numerical results presented in subsequent sections are obtained in an out-of-sample fashion by using exclusively a test set of $100,\!000$ additional simulated paths. 
	
	\subsection{Sensitivity analysis to risk measures}
	\label{subsec:sens_analysis}
	This section studies equal risk price values obtained under the semi-$\mathbb{L}^p$ and CVaR$_{\alpha}$ risk measures across different levels of risk aversion, i.e. different values for $p$ and $\alpha$. The main motivation is the following. \cite{carbonneau2021equal} observed that when hedging exclusively with the underlying stock, ERP under the CVaR$_{\alpha}$ measure produces option prices which are systematically inflated in comparison to those obtained under conventional risk-neutral measures, especially for OTM puts. This inflation phenomenon is significantly magnified with fat tails dynamics such as with a regime-switching (RS) model to an extent that can cast doubt on the applicability of ERP in practice. Furthermore, while the latter paper observed a positive relation between the risk aversion level $\alpha$ and equal risk prices $C_0^{\star}$, as shown in subsequent sections of this present paper, using smaller values for $\alpha$ leads to trading policies exhibiting poor risk mitigation performance with speculative behavior magnifying tail risk. 
	Consequently, the main motivation of this present section is to assess if the use of the semi-$\mathbb{L}^{p}$ class of risk measures helps alleviating this price inflation phenomenon while simultaneously resulting in optimized trading policies providing effective risk mitigation. Thus, a critical aspect of the sensitivity analysis performed in this section is the benchmarking of not only equal risk prices generated under different objective functions, but also the assessment of the effectiveness of the resulting global trading policies.
	
	\subsubsection{Regime-switching model}
	\label{subsubsec:RS_model}
	The conduction of a sensitivity analysis with respect to the objective function within the ERP framework necessitates the selection of a suitable dynamics for the underlying stock. Indeed, the model should incorporate salient stylized facts of financial markets with a specific focus on fat tails due to the assessment of the impact of objective functions within the ERP framework allowing more or less weights on extreme scenarios through their respective risk aversion parameter (i.e $\alpha$ and $p$ respectively for the CVaR$_{\alpha}$ and semi-$\mathbb{L}^{p}$ measures). Unless specified otherwise, this study considers a RS model for the risky asset dynamics. This class of model introduced in finance by \cite{hamilton1989new} exhibits, among others, fat tails, the leverage effect (i.e. negative correlation between assets returns and volatility) and heteroscedasticity. The examination of the impact of the presence of jump and volatility risk on $C_0^{\star}$ values generated with the semi-$\mathbb{L}^{p}$ objective functions is done in subsequent sections. Furthermore, unless specified otherwise, model parameters for the RS model (as well as for other dynamics considered subsequently) are estimated with maximum likelihood procedures on the same time series of daily log-returns on the S\&P 500 price index covering the period 1986-12-31 to 2010-04-01 (5863 observations). Parameter estimates are presented in \cref{appendix_MLE_Tables}.
	
	The description of the regime-switching model for the underlying stock is now formally defined. For $n=1,\ldots,N$, let $y_n \equiv \log(S_n^{(0,b)}/S_{n-1}^{(0,b)})$ be the time-$t_n$ log-return and $\{\epsilon_{n}\}_{n=1}^{N}$ be a sequence of independent and identically distributed (iid) standardized Gaussian random variables. The RS model assumes that the dynamics of the underlying stock changes between different regimes representing different economical states of the financial market. These regime changes are abrupt and they drastically impact the behavior of the dynamics of financial markets for a significant period of time, i.e. these regimes are persistent \citep{ang2012regime}. For instance, a two-regime RS model as considered in this study usually has a more bullish regime with positive expected returns and relatively small volatility, and a more bearish regime with negative expected returns and relatively large volatility. Prevalent examples of such regime changes are financial crises and important economical reforms.
	
	From a mathematical standpoint, the class of RS models characterizes regimes by an unobservable discrete-time Markov chain with a finite number of states, and models the conditional distribution of log-returns given the current regime as a Gaussian distribution with known parameters. More formally, denote the regimes as $\{h_n\}_{n=0}^{N}$ where $h_n \in \{1, \ldots, H\}$ is the regime in force during the time interval $[t_n, t_{n+1})$. The model specification for the transition probabilities of the Markov Chain can be stated as 
	\begin{align}
	\mathbb{P}(h_{n+1}=j|\mathcal{F}_n, h_n, \ldots, h_0) &= \gamma_{h_n, j}, \quad j = 1, \ldots, H, \label{eq:ref_transtion_matrix}
	\end{align}
	where $\Gamma \equiv \{\gamma_{i,j}\}_{i=1,j=1}^{H,H}$ is the transition matrix with $\gamma_{i,j}$ being the time-independent probability of moving from regime $i$ to regime $j$. Furthermore, the dynamics of log-returns have the representation
	\begin{align}
	y_{n+1} &= \mu_{h_n} \Delta + \sigma_{h_n} \sqrt{\Delta}\epsilon_{n+1}, \quad n = 0,\ldots, N-1,\nonumber
	\end{align}
	where $\{\mu_i, \sigma_i\}_{i=1}^{H}$ are model parameters representing the means and volatilities on a yearly basis of each regime. The use of a RS model entails that additional state variables related to the regimes must be added to feature vectors of neural networks through the vectors $\mathcal{I}_n$. Indeed, while regimes are unobservable, useful information can be filtered from the observed stock path prices. Let $\{\xi_n\}_{n=0}^{N}$ be the \textit{predictive probability process} where $\xi_n \equiv [\xi_{n,1},\ldots,\xi_{n,H}]$ and $\xi_{n,j} \equiv \mathbb{P}(h_n = j|\mathcal{F}_n)$. Under the RS model, $\mathcal{I}_n = \xi_n$ for $n=0,\ldots,N-1$. Following the work of \cite{franccois2014optimal}, the predictive probabilities can be computed recursively for $n=0,\ldots,N-1$ as
	$$\xi_{n+1,j} = \frac{\sum_{i=1}^{H}\gamma_{i,j} \phi_i(y_{n+1}) \xi_{n,i}}{\sum_{i=1}^{H}\phi_i(y_{n+1})\xi_{n,i}}, \quad j=1,\ldots,H,$$
	where $\phi_i$ is the probability density function of the Gaussian distribution with mean $\mu_i$ and volatility $\sigma_i$. For all numerical experiments, the time $0$ regime $h_0$ is sampled from the stationary distribution of the Markov Chain. Lastly, the benchmarking of equal risk prices to option prices obtained under conventional risk-neutral measures is also presented. Risk-neutral dynamics as well as the numerical scheme used to evaluate the risk-neutral price (including for alternative dynamics introduced subsequently) are presented in \cref{appendix:RN_dyn}.
	
	\subsubsection{Numerical results sensitivity analysis to objective function}
	\label{subsubsec:sensitivity_ERP_riskmeasure}
	\cref{table:sensitivity_analysis_ERP} presents equal risk prices obtained under the CVaR$_{\alpha}$ with $\alpha = 0.90, 0.95, 0.99$ as well as under the class of semi-$\mathbb{L}^{p}$ risk measures with $p = 2, 4, 6, 8, 10$. All equal risk prices are expressed relative to risk-neutral prices $C_0^{\mathbb{Q}}$. Hedging statistics obtained across the different objective functions are analyzed subsequently in \cref{subsubsec:hedgingstats_riskmeasure}. 	
	\begin{table}[ht]
		\caption {Sensitivity analysis of equal risk prices $C_0^{\star}$ for OTM ($K=90$), ATM ($K=100$) and ITM ($K=110$) put options of maturity $T=60/260$ under the regime-switching model.} \label{table:sensitivity_analysis_ERP}
		\renewcommand{\arraystretch}{1.15}
		\begin{adjustwidth}{-1in}{-1in} 
			\centering
			\begin{tabular}{ccccccccccc}
				\hline\noalign{\smallskip}
				& & \multicolumn{3}{c}{$C_0^{\star}$ under $\text{CVaR}_{\alpha}$} & & \multicolumn{5}{c}{$C_0^{\star}$  under semi-$\mathbb{L}^{p}$} \\
				\cline{3-5} \cline{7-11} $\text{Moneyness}$ & $C_0^{\mathbb{Q}}$ & $\text{CVaR}_{0.90}$ & $\text{CVaR}_{0.95}$ & $\text{CVaR}_{0.99}$ & & $\mathbb{L}^{2}$ & $\mathbb{L}^{4}$ & $\mathbb{L}^{6}$ & $\mathbb{L}^{8}$ & $\mathbb{L}^{10}$ \\
				\hline\noalign{\medskip} 
				$\text{OTM}$   & $0.56$  & $91\%$  & $119\%$ & $161\%$ & & $50\%$ & $88\%$  & $111\%$ & $140\%$  & $175\%$ \\
				$\text{ATM}$   & $3.27$  & $18\%$  & $24\%$  & $29\%$  & & $10\%$ & $17\%$  & $22\%$  & $28\%$   & $35\%$ \\
				$\text{ITM}$   & $10.36$ & $5\%$   & $7\%$   & $9\%$   & & $2\%$  & $5\%$   & $7\%$   & $8\%$    & $9\%$ \\
				\noalign{\medskip}\hline
			\end{tabular}%
		\end{adjustwidth}
		Notes: $C_0^{\star}$ results are computed based on $100,\!000$ independent paths generated from the regime-switching model under $\mathbb{P}$ (see \cref{subsubsec:RS_model} for model definition and \cref{appendix_MLE_Tables} for model parameters). Risk-neutral prices $C_0^{\mathbb{Q}}$ are computed under $\mathbb{Q}$-dynamics described in \cref{appendix:RN_dyn}. The training of neural networks is performed as described in \cref{subsec:calibration_NNETS} with hyperparameters presented in \cref{subsec:market_setup}. $C_0^{\star}$ are expressed relative to $C_0^{\mathbb{Q}}$ (\% increase).
	\end{table}
 
	Values from \cref{table:sensitivity_analysis_ERP} indicate that equal risk prices generated by the class of semi-$\mathbb{L}^{p}$ risk measures can span much more than the interval of prices obtained under the $\text{CVaR}_{\alpha}$ risk measures with the selected values for the confidence level $\alpha$. The latter observation holds across all moneyness levels for puts. For instance, the relative increase in the equal risk price $C_0^{\star}$ as compared to the risk-neutral price $C_0^{\mathbb{Q}}$ for OTM puts is $91\%, 119\%$ and $161\%$ under CVaR$_{0.90}$, CVaR$_{0.95}$ and CVaR$_{0.99}$, and ranges between $50\%$ to $175\%$ using the semi-$\mathbb{L}^{p}$ with $p$ going from $2$ to $10$. Similar observations can be made for ATM and ITM moneyness levels. Furthermore, the use of the semi-$\mathbb{L}^{2}$ risk measure entails a significant reduction of $C_0^{\star}$ as compared to the price obtained under the $\text{CVaR}_{0.90}$. Indeed, the relative increase in the equal risk price $C_0^{\star}$ with $p=2$ as compared to the risk-neutral price $C_0^{\mathbb{Q}}$ for OTM, ATM and ITM moneyness levels is respectively $50\%$, $10\%$ and $2\%$, which is significantly smaller than the corresponding relative increases of $91\%, 18\%$ and $5\%$ under the $\text{CVaR}_{0.90}$ measure. 
	Moreover, as expected, equal risk prices $C_0^{\star}$ generated with the class of semi-$\mathbb{L}^{p}$ risk measures show a positive relation with the risk aversion parameter $p$. This observation can be explained by a rationale analogous to that mentioned in \cite{carbonneau2021equal} under the CVaR$_{\alpha}$ risk measure case: since the put option payoff is bounded below at zero, the short position hedging error has a thicker right tail than the corresponding right tail of the long position hedging error. Consequently, an increase in the risk aversion parameter $p$ entails placing more weight on extreme hedging losses, which results in a larger increase of perceived residual risk exposure for the short position than for the long position. The latter entails that $C_0^{\star}$ must be increased to equalize the residual hedging risk of both parties. In conclusion, all these results clearly demonstrate the benefit of using the class of semi-$\mathbb{L}^{p}$ risk measures from the standpoint of pricing derivatives by not only spanning wider ranges of prices than these generated by the CVaR with conventional confidence levels, but by also significantly alleviating the inflated option prices phenomenon observed under the CVaR$_{\alpha}$. However, the question about whether or not the optimized global policies under the semi-$\mathbb{L}^{p}$ risk measures are effective from the standpoint of risk mitigation remains. This is examined in the following section.
	
	\subsubsection{Hedging performance benchmarking}
	\label{subsubsec:hedgingstats_riskmeasure}
	This section conducts the benchmarking of the neural networks trading policies hedging performance under the CVaR$_{\alpha}$ and semi-$\mathbb{L}^{p}$ objective functions. For the sake of brevity, hedging metrics values considered to compare the different policies are only presented for the short position hedge of the ATM put with the usual market setup, i.e. time-to-maturity of $T=60/260$ under the regime-switching model with daily stock hedges. \cref{table:RS_hedge_stats} presents hedging statistics of the global hedging policies obtained with the CVaR$_{\alpha}$ and semi-$\mathbb{L}^{p}$ risk measures with the same objective functions used to generate the $C_0^{\star}$ values in the previous section (i.e. $\alpha = 0.90, 0.95, 0.99$ and $p=2,4,6,8,10$). To compare the trading policies on common grounds, the initial portfolio value is set as the risk-neutral price with $V_0 = 3.27$ for all examples.\footnote{
	Recall that optimal policies under the CVaR$_{\alpha}$ risk measures are independent of $V_0$ due to the translation invariance property. Furthermore, the optimal policies obtained under the semi-$\mathbb{L}^{p}$ risk measures can be used not only with a specific value for $V_0$, but with an interval of initial capital investments that includes the risk-neutral price due to the proposed modified training algorithm in this paper.
	} Furthermore, hedging metrics used for the benchmarking consist of the VaR$_{\alpha}$ and CVaR$_{\alpha}$ statistics over various $\alpha$'s, the mean hedging error, the SMSE (i.e. semi-$\mathbb{L}^{2}$ metric) and the mean-squared-error (MSE). Note that all hedging statistics are estimated in an out-of-sample fashion on the test set of $100,\!000$ additional independent simulated paths.

	\begin{table}[ht]
	\caption {Hedging statistics for short position ATM put option of maturity $T=60/260$ under the regime-switching model.} \label{table:RS_hedge_stats}
	\renewcommand{\arraystretch}{1.15}
	\begin{adjustwidth}{-1in}{-1in} 
		\centering
		\begin{tabular}{lccccccccc}
			\hline\noalign{\smallskip}
			& \multicolumn{3}{c}{$\text{CVaR}_{\alpha}$} & & \multicolumn{5}{c}{\text{semi}-$\mathbb{L}^p$} \\
			\cline{2-4} \cline{6-10} Penalty & $\text{CVaR}_{0.90}$ & $\text{CVaR}_{0.95}$ & $\text{CVaR}_{0.99}$  & &  $\mathbb{L}^{2}$ & $\mathbb{L}^{4}$ & $\mathbb{L}^{6}$ & $\mathbb{L}^{8}$ & $\mathbb{L}^{10}$\\
			\hline\noalign{\medskip} 
			$\underline{Statistics}$   & & & & & & & & & \\
			$\text{Mean}$           & $0.11$ & $0.13$ & $0.14$                      & & $\textBF{-0.04}$ & $0.03$ & $0.11$ & $0.13$ & $0.15$ \\
			$\text{CVaR}_{0.90}$    & $\textBF{2.64}$   & $5.3\%$     & $22.6\%$    & & $5.4\%$      & $5.6\%$    & $7.4\%$   & $11.6\%$ & $16.9\%$ \\
			$\text{CVaR}_{0.95}$    & $3.41$   & $-\textBF{8.4\%}$    & $1.6\%$     & & $-1.6\%$     & $-5.1\%$   & $-5.8\%$  & $-3.6\%$ & $0.2\%$\\
			$\text{CVaR}_{0.99}$    & $6.86$   & $-31.7\%$   & $-\textBF{44.5\%}$   & & $-31.4\%$    & $-39.1\%$  & $-41.8\%$ & $-43.1\%$ & $-42.2\%$\\
			$\text{CVaR}_{0.999}$   & $19.99$  & $-48.5\%$   & $-76.1\%$            & & $-65.6\%$    & $-72.4\%$  & $-74.2\%$ & $-75.8\%$ & $-\textBF{76.4}\%$\\
			$\text{VaR}_{0.90}$     & $\textBF{1.75}$   & $34.7\%$    & $59.9\%$    & & $12.8\%$     & $21.3\%$   & $30.2\%$  & $36.7\%$  & $45.1\%$\\
			$\text{VaR}_{0.95}$     & $\textBF{2.08}$   & $21.9\%$    & $54.6\%$    & & $21.4\%$     & $25.9\%$   & $29.6\%$  & $37.6\%$  & $44.9\%$\\
			$\text{VaR}_{0.99}$     & $3.67$   & $-\textBF{9.6\%}$   & $-2.9\%$     & & $5.1\%$      & $-1.8\%$   & $-4.1\%$  & $-3.9\%$  & $-0.6\%$\\
			$\text{VaR}_{0.999}$    & $11.00$   & $-43.3\%$   & $-\textBF{62.5}\%$  & & $-47.6\%$    & $-55.4\%$  & $-57.8\%$ & $-60.3\%$ & $-60.4\%$ \\	
			$\text{SMSE}$           & $1.83$   & $-7.0\%$   & $6.8\%$               & & $-\textBF{33.5\%}$   & $-30.5\%$  & $-22.2\%$ & $-15.4\%$ & $-5.7\%$ \\
			$\text{MSE}$            & $2.93$   & $-1.8\%$    & $12.2\%$             & & $-\textBF{26.4\%}$   & $-24.2\%$  & $-15.6\%$ & $-9.7\%$   & $-0.2\%$\\
			\noalign{\medskip}\hline
		\end{tabular}%
	\end{adjustwidth}
	Notes: Hedging statistics are computed based on $100,\!000$ independent paths generated from the regime-switching model under $\mathbb{P}$ (see \cref{subsubsec:RS_model} for model definition and \cref{appendix_MLE_Tables} for model parameters). The training of neural networks is performed as described in \cref{subsec:calibration_NNETS} with hyperparameters presented in \cref{subsec:market_setup}. All hedging statistics except the mean hedging error are expressed relative to values obtained under the $\text{CVaR}_{0.90}$ penalty (\% increase). \textbf{Bold} values are the lowest across all penalties.
	\end{table}

	Hedging metrics values show that while the trading policy optimized with the CVaR$_{0.90}$ objective function entails the smallest values for CVaR$_{0.90}$, VaR$_{0.90}$ and VaR$_{0.95}$ statistics, it exhibits poor mitigation of tail risk as compared to the other policies. For instance, the relative reduction of the CVaR$_{0.99}$ statistic achieved with all other penalties than the CVaR$_{0.90}$ ranges between $31.4\%$ and $44.5\%$ as compared to the CVaR$_{0.90}$ trading policy. Similar observations can be made for the CVaR$_{0.999}$ and  VaR$_{0.999}$ statistics capturing extreme scenarios. The latter results cast doubt on the practical effectiveness of the CVaR$_{0.90}$ hedging policy from a risk mitigation standpoint, and thus also of trading policies optimized with CVaR$_\alpha$ with lower values for $\alpha$, due to their poor mitigation of risk for quantiles above the CVaR confidence level. This conclusion has important implications in the context of the ERP framework. Indeed, as shown in \cite{carbonneau2021equal}, the equal risk price $C_0^{\star}$ obtained with the CVaR$_{\alpha}$ exhibits a positive relation to $\alpha$ values. Consequently, the inflated equal risk price phenomenon observed under the class of CVaR$_{\alpha}$ measures cannot be effectively alleviated through the reduction of $\alpha$ as the resulting trading policies quickly exhibit poor hedging performance. On the other hand, hedging statistics obtained with the class of semi-$\mathbb{L}^{p}$ risk measures indicate that across all levels of risk aversion $p$ considered, optimized trading policies are effective for mitigating hedging risk. Recall that $p$ controls the weight associated with extreme hedging losses. From the combination of these hedging statistics values as well as equal risk price values presented in \cref{table:sensitivity_analysis_ERP}, we can conclude that the class of semi-$\mathbb{L}^{p}$ risk measures is a successful choice within the ERP framework by simultaneously generating lower and more reasonable equal risk prices than these obtained with the CVaR$_{\alpha}$ and by resulting in effective trading policies.
	

	\subsection{Sensitivity analysis to dynamics of risky assets}
	\label{subsec:diff_dyn}
	This section performs a sensitivity analysis of equal risk prices across different dynamics for the financial market. The motivation is to assess if the conclusion that the class of semi-$\mathbb{L}^{p}$ risk measures can dampen the inflated equal risk prices phenomenon as well as span wider price intervals than these obtained under the CVaR$_{\alpha}$ measures is robust to the presence of different equity risk features. For such a purpose, this paper considers the presence of jump risk with the Merton jump-diffusion model (MJD, \cite{merton1976option}) and of volatility risk with the GJR-GARCH model \citep{glosten1993relation}. The \cite{black1973pricing} and \cite{merton1973theory} (BSM) model is also considered due to its popularity and the fact that contrarily to the other dynamics, the BSM model does not exhibit fat tails. 
	The assessment of the impact of the choice of risk measure controlling the weight associated to extreme scenarios is thus also of interest under the BSM dynamics since the optimal hedging strategies, and thus equal risk prices, should be less sensitive to the risk aversion parameter under a dynamics without fat tails. 
	
	The dynamics of all three models is now formally presented. All model parameters are estimated with the same time series of daily log-returns on the S\&P 500 index covering the period 1986-12-31 to 2010-04-01 (5863 log-returns). Parameter estimates are presented in \cref{appendix_MLE_Tables}.

	\subsubsection{Black-Scholes model}
	The Black-Scholes model assumes that log-returns are iid Gaussian random variables of yearly mean $\mu - \sigma^2/2$ and volatility $\sigma$:
	$$y_{n} = \left(\mu - \frac{\sigma^2}{2}\right) \Delta + \sigma \sqrt{\Delta} \epsilon_n, \quad n = 1, \ldots, N.$$
	Stock prices have the Markov property under $\mathbb{P}$ with respect to the market filtration $\mathbb{F}$. The latter entails that no additional information should be added to the state variables of the neural networks, i.e. $\mathcal{I}_n = 0$ for all $n$.
	
	\subsubsection{GJR-GARCH model}
	The GJR-GARCH model relaxes the constant volatility assumption of the BSM model by assuming the presence of stochastic volatility which incorporates the leverage effect. Log-returns under this model have the representation
	\begin{align}
	y_n &= \mu + \sigma_n \epsilon_n, \nonumber
	\\ \sigma_{n+1}^{2} &= \omega + \upsilon \sigma_n^2(|\epsilon_n| - \gamma \epsilon_n)^2 + \beta \sigma_n^{2}, \nonumber
	\end{align}
	where $\{\sigma_n^{2}\}_{n=1}^{N+1}$ are the daily variances of log-returns, $\{\mu, \omega, \upsilon, \gamma, \beta\}$ are the model parameters with $\{\omega, \upsilon, \beta\}$ being positive real values and $\{\mu, \gamma\}$ real values. Note that given $\sigma_1^{2}$, 
	the sequence of variances $\sigma^2_2, \ldots, \sigma^2_{N+1}$ can be computed recursively with the observed path of log-returns. In this paper, the initial value $\sigma_1^{2}$ is set as the stationary variance of the process: $\sigma_1^{2} \equiv \E[\sigma_n^{2}] = \frac{\omega}{1 - \upsilon(1+\gamma^{2})-\beta}$. Furthermore, it can be shown that $\{S_n^{(0,b)}, \sigma_{n+1}\}_{n=0}^{N}$ is an $(\mathbb{F}, \mathbb{P})$-Markov bivariate process. Consequently, the periodic volatility is  added to the states variables of the neural networks at each time step: $\mathcal{I}_n = \sigma_{n+1}$ for $n=0,\ldots,N-1$.
	
	\subsubsection{Merton jump-diffusion model}
	\label{subsubsec:MJD_dyn}
	Contrarily to the GJR-GARCH model, the MJD dynamics assumes constant volatility, but deviates from the BSM assumptions by incorporating random Gaussian jumps to stock returns. Let $\{N_n\}_{n=0}^{N}$ be realizations of a Poisson process of parameter $\lambda > 0$, where $N_n$ represents the cumulative number of jumps of the stock price from time $0$ to time $t_n$. The \cite{merton1976option} model assumes that jumps, denoted by $\{\zeta_j\}_{j=1}^{\infty}$, are iid Gaussian random variables of mean $\mu_J$ and variance $\sigma_J^{2}$ under the physical measure:\footnote{
	The convention that $\sum_{j=N_{n-1}+1}^{N_n} \zeta_j = 0$ if $N_{n-1} = N_n$ is adopted.
	}
	\begin{align}
	y_n = \left(\nu - \lambda(e^{\mu_J + \sigma_J^2/2} -1) - \frac{\sigma^{2}}{2}\right)\Delta + \sigma \sqrt{\Delta}\epsilon_n + \sum_{j=N_{n-1}+1}^{N_n} \zeta_j, \nonumber
	\end{align}
	where $\{\epsilon_n\}_{n=1}^{N}$, $\{N_n\}_{n=0}^{N}$ and $\{\zeta_j\}_{j=1}^{\infty}$ are independent. Model parameters consist of  $\{\nu, \lambda, \sigma, \mu_J, \sigma_J\}$ where $\nu \in \mathbb{R}$ is the drift parameter and $\sigma > 0$ is the constant volatility term. Since stock returns are iid, this dynamics does not necessitate the addition of other state variables to the feature vectors, i.e. $\mathcal{I}_n = 0$ for all $n$.
	
	\subsubsection{Numerical results sensitivity analysis to dynamics}
	\cref{table:sensitivity_analysis_dyn_ERP} presents the sensitivity analysis of equal risk prices with the same setup as in previous sections, i.e. for put options of maturity $T=60/260$ with daily stock hedges, for the BSM, MJD and GJR-GARCH models. To save space, results are only presented for the OTM moneyness as the main conclusions are shared for both ATM and ITM moneyness levels. Furthermore, both the CVaR$_{\alpha}$ and semi-$\mathbb{L}^{p}$ classes of risk measures are considered with $\alpha = 0.90, 0.95, 0.99$ and $p=2,4,6,8,10$. 
	\begin{table}[ht]
		\caption {Sensitivity analysis of equal risk prices for OTM put options of maturity $T=60/260$ under the BSM, MJD and GJR-GARCH models.} \label{table:sensitivity_analysis_dyn_ERP} 
		\renewcommand{\arraystretch}{1.15}
		\begin{adjustwidth}{-1in}{-1in} 
			\centering
			\begin{tabular}{ccccccccccc}
				\hline\noalign{\smallskip}
				& & \multicolumn{3}{c}{$C_0^{\star}$ under $\text{CVaR}_{\alpha}$} & & \multicolumn{5}{c}{$C_0^{\star}$  under semi-$\mathbb{L}^{p}$} \\
				\cline{3-5} \cline{7-11} $\text{Dynamics}$ & $C_0^{\mathbb{Q}}$ & $\text{CVaR}_{0.90}$ & $\text{CVaR}_{0.95}$ & $\text{CVaR}_{0.99}$ & & $\mathbb{L}^{2}$ & $\mathbb{L}^{4}$ & $\mathbb{L}^{6}$  & $\mathbb{L}^{8}$ & $\mathbb{L}^{10}$ \\
				\hline\noalign{\medskip} 
				$\text{BSM}$         & $0.53$  & $5\%$   & $10\%$  & $17\%$  & & $3\%$  & $10\%$   & $22\%$  & $31\%$    & $43\%$ \\
				$\text{MJD}$         & $0.46$  & $23\%$  & $34\%$  & $129\%$ & & $15\%$ & $41\%$   & $71\%$  & $102\%$    & $125\%$ \\
				$\text{GJR-GARCH}$   & $0.57$  & $52\%$  & $71\%$  & $139\%$ & & $29\%$ & $96\%$   & $156\%$ & $219\%$   & $265\%$ \\
				\noalign{\medskip}\hline
			\end{tabular}%
		\end{adjustwidth}
		Notes: Equal risk prices $C_0^{\star}$ results are computed based on $100,\!000$ independent paths generated from the BSM, MJD and GJR-GARCH model under $\mathbb{P}$ (see \cref{subsec:diff_dyn} for models definitions under $\mathbb{P}$ and \cref{appendix_MLE_Tables} for model parameters). Risk-neutral prices $C_0^{\mathbb{Q}}$ are computed under $\mathbb{Q}$-dynamics described in \cref{appendix:RN_dyn}. The training of feedforward neural networks is performed as described in \cref{subsec:calibration_NNETS} with hyperparameters presented in \cref{subsec:market_setup}. $C_0^{\star}$ are expressed relative to $C_0^{\mathbb{Q}}$ (\% increase).
	\end{table}
	
	These results clearly demonstrate that the conclusion that equal risk prices generated by the class of semi-$\mathbb{L}^{p}$ risk measures can alleviate the price inflation phenomenon observed under the CVaR$_{\alpha}$ measures is robust to different dynamics. Indeed, by using the semi-$\mathbb{L}^{2}$ risk measure, OTM equal risk prices $C_0^{\star}$ exhibit a relative increase over risk-neutral prices $C_0^{\mathbb{Q}}$ of respectively $3\%, 15\%$ and $29\%$ under the BSM, MJD and GARCH models as compared to $5\%$, $23\%$ and $52\%$ under the CVaR$_{0.90}$ objective function. Furthermore, values presented in \cref{table:sensitivity_analysis_dyn_ERP} demonstrate that the observation made in the previous section under the RS model with respect to the fact that equal risk prices generated by the class of semi-$\mathbb{L}^{p}$ risk measures can span a large interval of prices which encompasses values obtained with the $\text{CVaR}_{\alpha}$ measures is robust to different dynamics of the financial markets. Lastly, it is interesting to observe that the length of the price intervals generated by both classes of risk measures varies significantly with the dynamics of the financial market. Indeed, under the BSM model, the relative increase of $C_0^{\star}$ as compared to $C_0^{\mathbb{Q}}$ ranges between $5\%$ to $17\%$ under the CVaR$_{\alpha}$ and between $3\%$ to $43\%$ under the semi-$\mathbb{L}^{p}$. On the other hand, with the GJR-GARCH dynamics, the relative increase in $C_0^{\star}$ under the CVaR$_{\alpha}$ ranges between $52\%$ to $139\%$, while under the semi-$\mathbb{L}^{p}$, it ranges between $29\%$ to $265\%$. Similar observations can be made under the MJD dynamics. This can be explained by the fact that contrarily to the other models, the BSM dynamics does not exhibits fat tails as the market incompleteness solely stems from discrete-time trading. Consequently, the trading policies are much less sensitive to the choice of risk aversion parameter $p$ or $\alpha$ under the BSM model, which results in equal risk price values that are less sensitive to risk aversion parameters.
	From these results, we can conclude that the choice of both the risky asset dynamics and of the risk measure among the classes of CVaR$_{\alpha}$ and semi-$\mathbb{L}^{p}$ measures has a material impact on equal risk prices, and this impact becomes more important as the dynamics exhibits fatter tails for risky assets returns.
	

	\subsection{Long-term maturity ERP with option hedges}
	\label{subsec:option_hedges}	
	This section examines the use of semi-$\mathbb{L}^{p}$ risk measures within the ERP framework for pricing long-term options with trades involving exclusively the underlying stock as compared to equal risk prices generated under the CVaR$_{\alpha}$ with trades involving shorter-term options. The motivation for this experiment is the following. The main finding of \cite{carbonneau2021deep} is that under the CVaR$_{\alpha}$ measure, hedging long-term puts with shorter-term options in the presence of jump or volatility risks significantly reduces equal risk prices as compared to trading exclusively the underlying stock. However, the expected trading cost of setting up a dynamic trading strategy based solely on option hedges can be impractical in some cases in the face of highly illiquid options. In such context, the hedger could potentially be restricted to a trading strategy relying exclusively on the underlying stock, which as shown in previous sections can inflate equal risk prices under the CVaR$_{\alpha}$ measure. The objective of this last section is thus to assess if the use of the semi-$\mathbb{L}^{p}$ risk measure can achieve a similar equal risk prices reduction when trading exclusively the underlying stock to that obtained when trading options with the CVaR$_{\alpha}$ objective function. The setup to perform this experiment is the same as the one considered in \cite{carbonneau2021deep}, and numerical values for equal risk prices generated with trades involving exclusively options under the CVaR$_{\alpha}$ are taken directly from the latter work. This setup is now recalled.
	
	The derivative to price and hedge is a 1-year put with $252$ days per year of moneyness levels OTM, ATM and ITM with strike prices of $90, 100$ and $110$, respectively. The annualized continuous risk-free rate is $r=0.03$. Also, as noted in \cite{carbonneau2021deep}, option trading strategies optimized with the confidence level $\alpha$ smaller than $0.95$ when using the CVaR as the objective function often leads to hedging strategies exhibiting poor tail risk mitigation. Thus, the convex risk measure considered as the benchmark in the present study is the CVaR$_{0.95}$ measure with trades involving either exclusively the underlying stock on a daily or monthly basis (i.e. $N=252$ or $N=12$, respectively), or by trading solely with ATM 1-month and 3-months calls and puts (i.e. $N=12$ or $N=4$, respectively). Following the work of \cite{carbonneau2021deep}, the pricing of options used as hedging instruments is done through the modeling of the daily variations of the ATM logarithm implied volatility dynamics under $\mathbb{P}$ as an autoregressive (AR) model of order $1$, named log-AR(1) hereafter. Furthermore, the model assumes for convenience that the ATM 1-month and 3-months implied volatilities are the same.\footnote{
	Note that traded options with different maturities are never used simultaneously in the same hedging simulation.
	} It is worth highlighting that the implied volatility model is used exclusively for pricing options used as hedging instruments, not for the 1-year put $\Phi$ to be priced. Also, note that while the rebalancing frequency is either daily, monthly or quarterly, IV variations are always generated on a daily basis.
	
	The log-AR(1) model is now formally defined. Denote by $\{IV_{n}\}_{n=0}^{252}$  the daily implied volatilities for the ATM calls and puts of $1$-month and $3$-months maturities which are used as hedging instruments. Also, let $\{Z_n\}_{n=1}^{252}$ be an additional sequence of iid standardized Gaussian random variables representing the random innovations of the log-IV dynamics. To capture the well-known leverage effect between asset returns and implied volatility variations (see for instance \cite{cont2002dynamics}), a correlation factor $\varrho \equiv corr(\epsilon_n, Z_n)$ set at $-0.6$ is considered where $\{\epsilon_n\}_{n=1}^{252}$ are the daily random innovations associated with stock returns. The log-AR(1) model has the represensation
	\begin{align}
	\log IV_{n+1} &= \log IV_{n} + \kappa(\vartheta  - \log IV_{n}) + \sigma_{IV} Z_{n+1}, \quad n = 0,\ldots, 251,\label{eq:ref_IV_model}
	\end{align}
	where $\{\kappa, \vartheta, \sigma_{IV}\}$ are the model parameters with $\kappa$ and $\vartheta$ as real values and $\sigma_{IV} > 0$. The initial value of the process is set at the long-term parameter with $\log IV_{0} \equiv \vartheta$. Moreover, the pricing of the calls and puts used as hedging instruments is performed with the well-known Black-Scholes formula with the annualized volatility set at the current implied volatility value. More precisely, denote by $C(IV, \Delta T, S, K)$ and $P(IV, \Delta T, S, K)$ the price of a call and put option respectively if the current implied volatility is $IV$, the time-to-maturity is $\Delta T$, the underlying stock price is $S$ and the strike price is $K$:
	\begin{align}
	C(IV, \Delta T, S, K) &\equiv S \mathcal{N}(d_1) - e^{-r\Delta T}K \mathcal{N}(d_2), \label{eq:ref_BSM_optprice_call}
	\\ P(IV, \Delta T, S, K)&\equiv e^{-r\Delta T} K \mathcal{N}(-d_2) - S \mathcal{N}(-d_1), \label{eq:ref_BSM_optprice_put}
	\end{align}
	where $\mathcal{N}(\cdot)$ denotes the cumulative distribution function of a standardized Gaussian random variable with 
	$$d_1 \equiv \frac{\log(\frac{S}{K}) + (r+\frac{IV^2}{2})\Delta T}{IV\sqrt{\Delta T}}, \quad d_2 \equiv d_1 - IV\sqrt{\Delta T}.$$
	Also, note that when option hedges are considered, the current implied volatility is added to the feature vectors of the neural networks. For instance, with $1$-month calls and puts hedges, the $n$th trade at time $t_n = n/12$ uses as input vectors for the neural networks $X_n = [S_{21 \times n}^{(0,b)}, IV_{21 \times n}, T- t_n, \mathcal{I}_{21 \times n}]$ for $n=0,1,\ldots,11$ where $21$ represents the number of days in a given month.\footnote{
	Note that with option hedges, the implied volatility of the options used as hedging instruments is added to feature vectors, not the price of each asset. This has the benefit of necessitating  one less state variable with the implied volatility instead of adding two state variables with the price of the call and put used for hedging. Furthermore, this is a reasonable choice from a theoretical standpoint as implied volatilities are simply a nonlinear transformation of options prices due to the bijection relation between the two values. 
	}

	Moreover, the dynamics of the underlying asset returns considered for this last section is once again the MJD dynamics, but with different parameters than in previous sections since the ones considered in \cite{carbonneau2021deep} are used for comparability purposes. The MJD as well as the log-AR(1) model parameters values are presented in \cref{table:MJD_with_options} and \cref{table:all_OU_params}. These parameters were chosen in an ad hoc fashion so as to produce reasonable values for the dynamics of the financial market.
	
	\begin{table}[ht]
		\caption {Parameters of the $1$-year Merton jump-diffusion model.} \label{table:MJD_with_options}
		\begin{adjustwidth}{-1in}{-1in} 
			\centering  
			\begin{tabular}{ccccc}
				\hline
				$\nu$ & $\sigma$ & $\lambda$ & $\mu_J$ & $\sigma_J$
				\\
				\hline\noalign{\medskip}
				$0.1111$  &  $0.1323$ & $0.25$ & $-0.10$ & $0.10$
				%
				\\    
				\noalign{\medskip}\hline
			\end{tabular}%
		\end{adjustwidth}
		\centering{Notes: $\nu$, $\sigma$ and $\lambda$ are on an annual basis.}
	\end{table}
	\begin{table} [ht]
		\caption {Parameters of the log-AR(1) model for the evolution of implied volatilities.}
		\label{table:all_OU_params}
		\begin{adjustwidth}{-1in}{-1in} 
			\centering
			\begin{tabular}{cccc}
				\hline
				$\kappa$ & $\vartheta $ & $\sigma_{\text{IV}}$ & $\varrho$
				\\
				\hline\noalign{\medskip}
				$0.15$  &  $\log(0.15)$ & $0.06$ & $-0.6$
				\\    
				\noalign{\medskip}\hline
			\end{tabular}%
		\end{adjustwidth}
	\end{table}
	
	\subsubsection{Numerical results with option hedges}
	\cref{table:ERP_with_options} presents equal risk prices $C_0^{\star}$ under CVaR$_{0.95}$ measure with daily or monthly stock trades as well as with 1-month or 3-months ATM calls and puts trades. Note that the latter values are from Table $3$ of \cite{carbonneau2021deep}.\footnote{
	The type of neural networks considered in \cite{carbonneau2021deep} is the long-short term memory (LSTM). The current paper found that FFNN trading policies performed significantly better for the numerical experiments conducted under the semi-$\mathbb{L}^{p}$ risk measure which motivated their use over LSTMs. The reader is referred to Section $3$ of \cite{carbonneau2021deep} for the formal description of the LSTM architecture.
	} Furthermore, $C_0^{\star}$ values under the semi-$\mathbb{L}^{2}$ objective function with daily and monthly stock hedges are also presented.
	
	\begin{table}[ht]
		\caption {Sensitivity analysis of equal risk prices to jump risk for OTM ($K=90$), ATM ($K=100$) and ITM ($K=110$) put options of maturity $T=1$.} \label{table:ERP_with_options}
		\renewcommand{\arraystretch}{1.15}
		\begin{adjustwidth}{-1in}{-1in} 
			\centering
			\begin{tabular}{cccccccc}
				\hline\noalign{\smallskip}
				& \multicolumn{4}{c}{$C_0^{\star}$ under $\text{CVaR}_{0.95}$} & & \multicolumn{2}{c}{$C_0^{\star}$  under semi-$\mathbb{L}^{2}$} \\
				\cline{2-5} \cline{7-8} $\text{Moneyness}$ & $\text{Daily stock}$ & $\text{Monthly stock}$ & $\text{1-month opts}$ & $\text{3-months opts}$ & & $\text{Daily stock}$ & $\text{Monthly stock}$ \\
				\hline\noalign{\medskip} 
				$\text{OTM}$   & $2.58$  & $2.60$  & $2.24$  & $2.08$  & & $2.18$   & $2.23$   \\
				$\text{ATM}$   & $6.01$  & $5.77$  & $5.36$  & $5.12$  & & $5.38$   & $5.22$   \\
				$\text{ITM}$   & $11.68$ & $11.44$ & $10.86$ & $10.51$ & & $10.42$  & $10.54$  \\
				\noalign{\medskip}\hline
			\end{tabular}%
		\end{adjustwidth}
		Notes: These results are computed based on $100,\!000$ independent paths generated from the MJD model under $\mathbb{P}$ (see \cref{subsubsec:MJD_dyn} for model definition and \cref{table:MJD_with_options} for model parameters). Options used as hedging instruments are priced with implied volatility modeled with a log-AR(1) dynamics (see \cref{subsec:option_hedges} for model description and \cref{table:all_OU_params} for parameters values). Values for $C_0^{\star}$ under CVaR$_{0.95}$ are from Table $3$ of \cite{carbonneau2021deep}. Values for $C_0^{\star}$ under semi-$\mathbb{L}^{2}$ are obtained with the training algorithm described in \cref{subsubsec_nontrans_inv}.
	\end{table}
	Numerical results indicate that the use of the semi-$\mathbb{L}^{2}$ objective function is successful at reducing significantly equal risk prices when relying on trades involving exclusively the underlying stock. Indeed, the relative reduction in $C_0^{\star}$ obtained by using the semi-$\mathbb{L}^{2}$ risk measure as compared to the CVaR$_{0.95}$ for OTM, ATM and ITM moneyness levels is respectively $15\%, 11\%$ and $11\%$ with daily stock and $14\%, 10\%$ and $8\%$ with monthly stock rebalancing.\footnote{
	For instance, if $C_0^{\star}(\text{CVaR}_{0.95})$ and $C_0^{\star}(\mathbb{L}^{2})$ are respectively equal risk prices under the CVaR$_{0.95}$ and semi-$\mathbb{L}^{2}$ objective functions, the relative reduction is computed as $1 - \frac{C_0(\mathbb{L}^{2})}{C_0^{\star}(\text{CVaR}_{0.95})}.$
	} Furthermore, equal risk prices values under the semi-$\mathbb{L}^{2}$ risk measure with daily or monthly stock hedges are relatively close to those obtained with 1-month or 3-months option hedges under the CVaR$_{0.95}$. These results have important implications for ERP procedures. Indeed, this demonstrates that in the face of highly illiquid options, the use of the semi-$\mathbb{L}^{p}$ class of risk measures with stock hedges can effectively reduce equal risk prices to levels similar than these obtained with option hedges under the CVaR$_{\alpha}$ measures. This avenue is thus successful to alleviate the price inflation phenomenon when using ERP procedures for the pricing of long-term options. It is worth highlighting that in the presence of jump risk, the use of options as hedging instruments is much more effective for risk mitigation as compared to hedging strategies involving exclusively the underlying stock (see for instance \cite{coleman2007robustly} and \cite{carbonneau2021deepIME}). Nevertheless, $C_0^{\star}$ values presented in \cref{table:ERP_with_options} indicate that when setting up trading strategies with options is impractical due to high expected trading costs, the use of stock hedges coupled with semi-$\mathbb{L}^{p}$ risk measures can effectively reduce option prices.

	\section{Conclusion}
	\label{section:conclusion}

	This paper studies the class of semi-$\mathbb{L}^{p}$ risk measures in the context of equal risk pricing (ERP) for the valuation of European financial derivatives. The ERP framework prices contingent claims as the initial hedging portfolio value which equates the residual hedging risk of the long and short positions under optimal hedging strategies. Despite lacking the translation invariance property which complexifies the numerical evaluation of equal risk prices, the use of semi-$\mathbb{L}^{p}$ risk measures as the objective functions measuring residual hedging risk is shown to have several preferable properties over the use of the $\text{CVaR}_{\alpha}$, the latter being explored for instance in \cite{carbonneau2021equal} and \cite{carbonneau2021deep} in the context of ERP. The optimal hedging problems underlying the ERP framework are solved with deep reinforcement learning procedures by representing trading policies with neural networks as proposed in the work of \cite{buehler2019deep}. 
	A modification to the training algorithm for neural networks is presented in this current paper to tackle the additional complexity of using semi-$\mathbb{L}^{p}$ risk measures within the ERP framework. This modification consists in training the neural networks to learn the optimal mappings for an interval of initial capital investments instead of a unique fixed value. The latter is shown not to lead to material deterioration in the hedging accuracy of the neural networks trading policies. 
	
	Several numerical experiments are performed to examine option prices generated by the ERP framework under the class of semi-$\mathbb{L}^{p}$ risk measures. First, a sensitivity analysis of equal risk price values with respect to the choice of objective function is conducted by comparing prices obtained with the CVaR$_{\alpha}$ and semi-$\mathbb{L}^{p}$ objectives across different values of $\alpha$ and $p$ controlling the risk aversion of the hedger. 
	Numerical results demonstrate that equal risk prices under the semi-$\mathbb{L}^{p}$ risk measures are spanning a larger interval of values than the one obtained with the CVaR$_{\alpha}$, thereby allowing alleviating the price inflation phenomenon observed under the CVaR$_{\alpha}$ documented in previous studies. Furthermore, the trading policies parameterized as neural networks are shown to be highly effective for risk mitigation under the semi-$\mathbb{L}^{p}$ objective functions across all values of $p$ considered, with the risk aversion parameter controlling the relative weight associated with extreme scenarios. 
	Moreover, additional numerical experiments show that the use of the semi-$\mathbb{L}^{2}$ objective function for the pricing of long-term puts with hedges exclusively relying on the underlying asset is successful at reducing equal risk prices roughly to the level of prices produced with option hedges under the CVaR$_{\alpha}$ objective function. The latter conclusion is highly important in the context of ERP as it demonstrates that in the case where options are not or cannot be used within the hedging strategy, the ERP methodology used in conjunction with the semi-$\mathbb{L}^{p}$ class of risk measures can produce reasonable option prices.

	
	
	\section{Acknowledgements}
	Alexandre Carbonneau gratefully acknowledges financial support from the Fonds de recherche du Qu\'ebec - Nature et technologies (FRQNT, grant number 205683) and The Montreal Exchange. Fr{\'e}d{\'e}ric Godin gratefully acknowledges financial support from Natural Sciences and Engineering Research Council of Canada (NSERC, grant number RGPIN-2017-06837).
	

	\bibliographystyle{apalike}
	\bibliography{Biblio_ERP_SMSE_penalty}


	\appendix

	
	\section{Pseudo-code}
	\label{appendix:pseudo_code}
	This section presents the pseudo-codes for the training of neural networks and the bisection method. \cref{algo:pseudo_code_training} describes the pseudo-code to carry out a single SGD step, i.e. given $\theta_j$ and the initial portfolio value $V_0$, the steps to perform an update of the set of  parameters to $\theta_{j+1}$. Without loss of generality, the training pseudo-code is presented only for the short neural network $F_{\theta}^{\mathcal{(S)}}$ and for trades involving only the underlying stock. 
	The update rule for portfolio values in step (6) of \cref{algo:pseudo_code_training} can be obtained directly from the self-financing representation of $V_n^{\delta}$ as shown below
	\begin{align}
	V_n^{\delta} &= B_n(V_0^{\delta} + G_n^{\delta}) \nonumber
	\\ &= B_n\left(V_0^{\delta} + G_{n-1}^{\delta} + \delta_n^{(0:D)} \bigcdot (B_n^{-1}S_{n-1}^{(e)} - B_{n-1}^{-1}S_{n-1}^{(b)}) \right) \nonumber
	\\ &= \frac{B_n}{B_{n-1}} V_{n-1}^{\delta} + \delta_n^{(0:D)} \bigcdot (S_{n-1}^{(e)} - \frac{B_n}{B_{n-1}}S_{n-1}^{(b)})  \nonumber
	\\ &= e^{r \Delta} V_{n-1}^{\delta} + \delta_n^{(0:D)} \bigcdot (S_{n-1}^{(e)} - e^{r\Delta}S_{n-1}^{(b)}).  \label{eq:ref_Vn_update}
	\end{align}
	
	\begin{algorithm}
		\caption{Pseudo-code training neural networks $F_{\theta}^{(\mathcal{S})}$ with underlying stock hedges\\
			Input: $\theta_j, V_{0}^{\delta}$                              \\
			Output: $\theta_{j+1}$}
		\label{algo:pseudo_code_training}
		\begin{algorithmic}[1]
			\For {$i=1,\ldots,N_{\text{batch}}$}  \Comment{Loop over each path of minibatch}
			\State $X_{0,i} =[T, \log(S_{0,i}^{(0,b)}/K), V_{0,i}^{\delta}/\tilde{V}, \mathcal{I}_{0,i}]$ \Comment{Time-$0$ feature vector of $F_{\theta}^{(\mathcal{S})}$}
			\For {$n=0,\ldots,N-1$} 
			\State $\delta_{n+1,i}^{(0)} \leftarrow$ \text{time-$t_{n}$ output of FFNN $F_{\theta}^{(\mathcal{S})}$ with $\theta = \theta_j$} 		
			\State $S_{n+1,i}^{(0,b)} = S_{n,i}^{(0,b)} e^{y_{n+1,i}}$   \Comment{Sample next stock price}
			\State $V_{n+1,i}^{\delta} = e^{r\Delta}V_{n,i}^{\delta} + \delta_{n+1,i}^{(0)}(S_{n+1,i}^{(0,b)} - e^{r\Delta} S_{n,i}^{(0,b)})$ \Comment{See \eqref{eq:ref_Vn_update} for details}
			\State $\mathcal{I}_{n+1,i} \leftarrow$ \text{update additional state variables}
			\State $X_{n+1,i} =[T - t_n, \log(S_{n+1,i}^{(0,b)}/K), V_{n+1,i}^{\delta}/\tilde{V}, \mathcal{I}_{n+1,i}]$ \Comment{
				Time $t_{n+1}$ feature vector of $F_{\theta}^{\mathcal{(S)}}$}
			\EndFor
			\State $\Phi(S_{N,i}^{(0,b)})=\max(K-S_{N,i}^{(0,b)}, 0)$
			\State $\pi_{i,j}=\Phi(S_{N,i}^{(0,b)}) - V_{N,i}^{\delta}$
			\EndFor
			\State $\widehat{J} = \left(\frac{1}{N_{\text{batch}}}\sum_{i=1}^{N_{\text{batch}}} \pi_{i,j}^{p} \mathds{1}_{ \{\pi_{i,j}>0\} }\right)^{1/p}$
			\State $\eta_j \leftarrow$ Adam algorithm
			\State $\theta_{j+1} = \theta_{j} - \eta_{j} \nabla_{\theta} \widehat{J}$ \Comment{$\nabla_{\theta} \widehat{J}$ computed with Tensorflow}
	\end{algorithmic}
	Notes: Subscript $i$ represents the $i$th simulated path among the minibatch of size $N_{\text{batch}}.$ Also, the time-$0$ feature vector is fixed for all paths, i.e. $S_{0,i}^{(0,b)} = S_{0}^{(0,b)}$, $V_{0,i}^{\delta}=V_{0}^{\delta}$ and $\mathcal{I}_{0,i} = \mathcal{I}_{0}$.
	\end{algorithm}
	
 \cref{algo:pseudo_code_bisection} presents the pseudo-code for the bisection algorithm taking as inputs the two trained neural networks $F_{\theta}^{(\mathcal{L})}$ and $F_{\theta}^{(\mathcal{S})}$ as well as the initial search range $[V_A,V_B]$ so as to output the equal risk price.  
\begin{algorithm}
	\caption{Pseudo-code bisection algorithm\\
		Input: $F_{\theta}^{(\mathcal{L})}$ and $F_{\theta}^{(\mathcal{S})}$ trained neural networks, initial search range $[V_A, V_B]$ and test set paths\\
		Output: $C_0^{\star}$}
	\label{algo:pseudo_code_bisection}
	\begin{algorithmic}[1]
		\State nbs\_iter = 0, $\Delta(V) = \infty$  
		\While {$|\Delta(V)| > \zeta$ and nbs\_iter $<$ max\_iter}
		\State $V = 0.5(V_A + V_B)$  
		\State Compute $\tilde{\epsilon}^{(\mathcal{L})}(-V)$ and $\tilde{\epsilon}^{(\mathcal{S})}(V)$ on the test set with $F_{\theta}^{(\mathcal{L})}$ and $F_{\theta}^{(\mathcal{S})}$
		\State $\Delta(V) = \tilde{\epsilon}^{(S)}(V) - \tilde{\epsilon}^{(L)}(-V)$ 
		\If {$\Delta(V) > 0$}              
		\State $V_A \leftarrow V$
		\Else
		\State $V_B \leftarrow V$
		\EndIf
		\State nbs\_iter $\leftarrow$ nbs\_iter + 1
		\EndWhile
		\State $C_0^{\star} = V$.
	\end{algorithmic}
	Notes: $\zeta$ and max\_iter represent respectively the admissible level of pricing error and the maximum number of iterations for the bisection algorithm. For all numerical experiments conducted in \cref{section:numerical_results}, $\zeta$ is set to $0.01$ and max\_iter to $100$.
\end{algorithm}


\section{Validation of modified training algorithm}
\label{appendix:proof_convergence_results}
The goal of this section is to demonstrate that the proposed modification to the training algorithm described in \cref{subsubsec_nontrans_inv} to tackle the non-translation invariant risk measures case of the ERP framework does not materially impact the optimized neural networks hedging performance. Denote by $F_\theta$ the neural network trained with the additional step of sampling $V_0 \in [V_A, V_B]$ on top of the minibatch of paths at the beginning of each stochastic gradient descent step. One conclusive test to validate that the proposed modification does not deteriorate the neural networks accuracy is to compare the hedging performance of $F_{\theta}$ assuming $V_0 = V^{\star}$ to another neural network denoted as $F_{\theta}^{\text{fixed}}$ trained exclusively with a fixed initial capital investment set at $V^{\star}$. If $F_{\theta}$ exhibits similar hedging performance to $F_{\theta}^{\text{fixed}}$ over multiple iterations of $V^{\star}$, this demonstrates that $F_{\theta}$ accurately learned the optimal trading policy over a range of possible initial capital investments. 

The experiment conducted to perform the latter test is now formally presented. The setup considered is similar to the one presented in \cref{subsec:market_setup} with the hedging of an ATM put option of maturity $T=60/260$ with daily stock hedges under the regime-switching model. The steps are the following for all semi-$\mathbb{L}^{p}$ objective functions, $p \in \{2,4,6,8,10\}$:
\begin{itemize}
	\item [1)] Train $F_{\theta}$ with the procedure described \cref{subsubsec_nontrans_inv} where $V_0$ is sampled in the interval $[0.75C_0^{\mathbb{Q}}, 1.50C_0^{\mathbb{Q}}]$ at the beginning of each SGD step with $C_0^{\mathbb{Q}}$ being the risk-neutral price. A total of $100$ epochs is used on the train set.
	\item [2)] For a fixed randomly sampled value $V^* \in [0.75C_0^{\mathbb{Q}}, 1.50C_0^{\mathbb{Q}}],$ set $V_0 = V^*$ and train $F_{\theta}^{\text{fixed}}$ with the methodology described in \cref{subsubsec:fixed_V0}. A total of three iterations of this step is performed (i.e. three different values of $V^{\star}$ are considered).
	\item [3)] For the three sampled values of $V^{\star}$, compute the semi-$\mathbb{L}^{p}$ statistics on the test set with $F_{\theta}$ and $F_{\theta}^{\text{fixed}}$.
\end{itemize}

\begin{table}[ht]
	\caption {Semi-$\mathbb{L}^{p}$ statistics of the modified training algorithm for ATM ($K=100$) put options of maturity $T=60/260$ under the regime-switching model.} \label{table:proof_of_convergence_RS}
	\renewcommand{\arraystretch}{1.15}
	\begin{adjustwidth}{-1in}{-1in} 
		\centering
		\begin{tabular}{ccccccccccccccc}
			\hline\noalign{\smallskip}
			& \multicolumn{2}{c}{$\mathbb{L}^{2}$} & & \multicolumn{2}{c}{$\mathbb{L}^{4}$} & & \multicolumn{2}{c}{$\mathbb{L}^{6}$} & & \multicolumn{2}{c}{$\mathbb{L}^{8}$} & & \multicolumn{2}{c}{$\mathbb{L}^{10}$} \\
			\cline{2-3} \cline{5-6} \cline{8-9} \cline{11-12} \cline{14-15}$V_0$ & $F_{\theta}$ & $F_{\theta}^{\text{fixed}}$ & & $F_{\theta}$ & $F_{\theta}^{\text{fixed}}$ & & $F_{\theta}$ & $F_{\theta}^{\text{fixed}}$ & & $F_{\theta}$ & $F_{\theta}^{\text{fixed}}$ & & $F_{\theta}$ & $F_{\theta}^{\text{fixed}}$ \\
			\hline\noalign{\medskip} 
			%
			$4.343$ & $0.6236$  & $0.6209$ & & $1.1523$  & $1.1483$  & & $1.5291$  & $1.5466$    & & $1.8981$   & $1.8628$ & & $2.2747$   & $2.2069$ \\
			$2.503$ & $1.5457$  & $1.5355$ & & $2.2485$  & $2.2466$  & & $2.6865$  & $2.6843$    & & $3.0126$   & $3.0277$ & & $3.3121$   & $3.3311$ \\
			$4.005$ & $0.7596$  & $0.7578$ & & $1.3288$  & $1.3317$  & & $1.716$   & $1.7292$    & & $2.0703$   & $2.0542$ & & $2.4308$   & $2.3866$ \\
			\noalign{\medskip}\hline
		\end{tabular}%
	\end{adjustwidth}
	Notes: semi-$\mathbb{L}^{p}$ statistics results are computed based on $100,\!000$ independent paths generated with the regime-switching model under $\mathbb{P}$ (see \cref{subsubsec:RS_model} for model definition and \cref{appendix_MLE_Tables} for model parameters). $F_\theta$ is the neural network trained with the modified algorithm described in \cref{subsubsec_nontrans_inv}. $F_{\theta}^{\text{fixed}}$ is the neural network trained with fixed initial capital investment of $V_0$ as described in \cref{subsubsec:fixed_V0}.  
\end{table}
\cref{table:proof_of_convergence_RS} presents the semi-$\mathbb{L}^{p}$ statistics for the three values of $V_0 = V^{\star}$ with $p=2,4,6,8,10$. These results clearly demonstrate that the modified training algorithm does not materially impact the accuracy of the neural network as the difference in semi-$\mathbb{L}^{p}$ statistics between the FFNNs $F_{\theta}$ and $F_{\theta}^{\text{fixed}}$ is most often marginal.


\section{Maximum likelihood estimates results}
\label{appendix_MLE_Tables}
This section presents maximum likelihood model parameters estimates for the different risky asset dynamics considered in numerical experiments of \cref{subsec:sens_analysis} and \cref{subsec:diff_dyn}. All parameters are estimated with the same time-series of daily log-returns on the S\&P 500 index for the period 1986-12-31 to 2010-04-01 (5863 log-returns). Estimated parameters are presented in \cref{table:MLE_SSP500_19861231_20100401_all_models} to \ref{table:MLE_SSP500_19861231_20100401_MJD}. 
	
	\begin{table}[!htbp]
	\caption {Maximum likelihood parameter estimates of the Black-Scholes model.} \label{table:MLE_SSP500_19861231_20100401_all_models}
	\begin{adjustwidth}{-1in}{-1in} 
		\centering
		\begin{tabular}{cc}
			\hline
			$\mu$ & $\sigma$
			\\
			\hline\noalign{\medskip}
			$0.0892$  &  $0.1952$
			\\    
			\noalign{\medskip}\hline
		\end{tabular}%
	\end{adjustwidth}
	\centering{Notes: Both $\mu$ and $\sigma$ are on an annual basis.}
	\vspace{1.1cm}
	
	\caption {Maximum likelihood parameter estimates of the GJR-GARCH model.}
	\begin{adjustwidth}{-1in}{-1in}
		\centering
		\begin{tabular}{ccccc}
			\hline
			$\mu$ & $\omega$ & $\upsilon$ & $\gamma$ & $\beta$
			\\
			\hline\noalign{\medskip}
			$2.871\text{e-}04$  &  $1.795\text{e-}06$ & $0.0540$ & $0.6028$ & $0.9105$
			\\    
			\noalign{\medskip}\hline
		\end{tabular}%
	\end{adjustwidth}
	
	\vspace{1.1cm}
	
	\caption {Maximum likelihood parameter estimates of the regime-switching model.}
	\begin{adjustwidth}{-1in}{-1in} 
		\centering
		\begin{tabular}{ccc}
			\hline
			& \multicolumn{2}{c}{$\text{Regime}$}
			\\			
			\hline
			$\text{Parameter}$ & $1$ & $2$
			\\ 
			$\mu$ & $0.1804$ & $-0.2682$
			\\    
			$\sigma$ & $0.1193$ & $0.3328$
			\\
			$\nu$ & $0.7543$ & $0.2457$
			\\
			\hline
			$\Gamma$ & $0.9886$ & $0.0114$
			\\
			& $0.0355$ & $0.9645$      
			\\    
			\noalign{\medskip}\hline
		\end{tabular}%
	\end{adjustwidth}
	Notes: Parameters were estimated with the EM algorithm of \cite{dempster1977maximum}. $\nu$ represent probabilities associated with the stationary distribution of the Markov chain. $\mu$ and $\sigma$ are on an annual basis. 
	
	\vspace{1.1cm}
	
	\caption {Maximum likelihood parameter estimates of the Merton jump-diffusion model.} \label{table:MLE_SSP500_19861231_20100401_MJD}
	\begin{adjustwidth}{-1in}{-1in} 
		\centering  
		\begin{tabular}{ccccc}
			\hline
			$\nu$ & $\sigma$ & $\lambda$ & $\mu_J$ & $\sigma_J$
			\\
			\hline\noalign{\medskip}
			$0.0875$  &  $0.1036$ & $92.3862$ & $-0.0015$ & $0.0160$
			\\    
			\noalign{\medskip}\hline
		\end{tabular}%
	\end{adjustwidth}
	\centering{Notes: $\nu$, $\sigma$ and $\lambda$ are on an annual basis.}
	
\end{table}

\section{Risk-neutral dynamics}
\label{appendix:RN_dyn}
This section presents the risk-neutral dynamics for the RS, BSM, GARCH and MJD models. The absence of arbitrage opportunities implied by each model entails by the first fundamental theorem of asset pricing that there exists a probability measure $\mathbb{Q}$ such that $\{S_n^{(0,b)}e^{-r t_n}\}_{n=0}^{N}$ is an $(\mathbb{F}, \mathbb{Q})$-martingale \citep{delbaen1994general}. Denote by $\{\epsilon_n^{\mathbb{Q}}\}_{n=1}^{N}$ a sequence of iid standardized Gaussian random variables under $\mathbb{Q}$. Hereby are the $\mathbb{Q}$-dynamics for the four models described in \cref{subsubsec:RS_model} and \cref{subsec:diff_dyn} as well as the corresponding methods to compute the risk-neutral price $C_0^{\mathbb{Q}}$ of European puts. 

\subsection{Regime-switching}
The change of measure used in this study is the popular choice of shifting the drift to obtain risk-neutrality and model invariance as considered for instance in \cite{hardy2001regime}. Under this change of measure $\mathbb{Q}$, the drift $\mu_i \Delta$ in each regime is shifted to $(r - \sigma_i^2/2)\Delta$, and the transition probabilities are left unchanged. The risk-neutral dynamics has the representation
\begin{align}
y_{n+1}  = \left(r - \frac{\sigma_{h_{n}}^{2}}{2}\right)\Delta + \sigma_{h_{n}} \sqrt{\Delta}\epsilon^{\mathbb{Q}}_{n+1}, \quad n = 0,\ldots,N-1. \nonumber
\end{align} 
To compute the risk-neutral price of $\Phi$, the approach used follows the work of \cite{godin2019option} (see Section $5.3$ of the latter paper). Let $\mathbb{H} \equiv \{\mathcal{H}_{n}\}_{n=0}^{N}$ be the filtration generated by the regimes and $\mathbb{G}$ be the filtration containing all latent factors and all market information available to financial participants, i.e. $\mathbb{G} \equiv \mathbb{F}  \vee \mathbb{G}$. Using the law of iterative expectations, the risk-neutral price of $\Phi$ has the representation
\begin{align}
C_0^{\mathbb{Q}} &\equiv e^{-rT}\E^{\mathbb{Q}}[\Phi(S_N^{(0,b)})|\mathcal{F}_{0}] \nonumber
\\ &= e^{-rT}\E^{\mathbb{Q}}\left[\E^{\mathbb{Q}}[\Phi(S_N^{(0,b)})|\mathcal{G}_0]|\mathcal{F}_{0}\right] \nonumber
\\ &= e^{-rT} \sum_{i=1}^{H} \xi_{0,i}^{\mathbb{Q}} \E^{\mathbb{Q}}[\Phi(S_N^{(0,b)})|h_0 = i, S_0^{(0,b)}], \label{eq:ref_price_RS}
\end{align}
where $\xi_{0,i}^{\mathbb{Q}}$ is assumed to be equal to $\xi_{0,i}^{\mathbb{P}}$ for all regimes $i$, i.e. to the stationary distribution of the Markov chain under $\mathbb{P}$. The computation of the conditional expectations in \eqref{eq:ref_price_RS} can be done for instance with Monte Carlo simulations or with the closed-form solution of \cite{hardy2001regime} when $H=2$.

\subsection{BSM}
The change of measure from $\mathbb{P}$ to $\mathbb{Q}$ under the BSM dynamics is the one obtained with the discrete-time version of the Girsanov theorem: there exists a market price of risk process denoted as $\psi \equiv \{\psi_n\}_{n=1}^{N}$ such that $\epsilon_n^{\mathbb{Q}} = \epsilon_n + \psi_n$. By setting $\psi_n \equiv \sqrt{\Delta}(\frac{\mu - r}{\sigma})$, it is easy to show that $\{S_n^{(0,b)} e^{-rt_n}\}_{n=0}^{N}$ is an $(\mathbb{F}, \mathbb{Q})$-martingale and that the $\mathbb{Q}$-dynamics of log-returns is
$$y_n = \left(r - \frac{\sigma^{2}}{2}\right)\Delta + \sigma \sqrt{\Delta} \epsilon_n^{\mathbb{Q}}.$$
Risk-neutral put option prices presented in this paper are computed with the well-known Black-Scholes closed-form solution.

\subsection{GARCH}
The change of measure from $\mathbb{P}$ to $\mathbb{Q}$ considered is the one from \cite{duan1995garch} where the one-period conditional expected log-return is shifted, but the one-period conditional variance is unchanged when going from the physical to the risk-neutral measure. More precisely, let $\epsilon_n^{\mathbb{Q}} = \epsilon_n + \psi_n$ where $\psi \equiv \{\psi_n\}_{n=1}^{N}$ is predictable with respect to the filtration $\mathbb{F}$. The one-period expected conditional gross return under $\mathbb{Q}$ must be equal to the one-period risk-free rate accrual factor for $n=1,\ldots,N$:
$$\E^{\mathbb{Q}}[e^{y_n}|\mathcal{F}_{n-1}] = \E^{\mathbb{Q}}[e^{\mu - \psi_n\sigma_n + \sigma_n \epsilon_n^{\mathbb{Q}}}|\mathcal{F}_{n-1}] = e^{\mu - \psi_n \sigma_n + \sigma_n^{2}/2} = e^{r \Delta}.$$
Thus, $\psi_n$ has the representation 
\begin{align}
\psi_n \equiv \frac{\mu - r \Delta + \sigma_n^{2}/2}{\sigma_n}, \quad n = 1,\ldots,N. \label{eq:ref_GARCH_risk}
\end{align}
With \eqref{eq:ref_GARCH_risk}, the GARCH risky asset dynamics under $\mathbb{Q}$ is
\begin{align}
y_n &= r \Delta - \sigma_n^{2}/2 + \sigma_n \epsilon_n^{\mathbb{Q}}, \nonumber
\\ \sigma_{n+1}^{2} &= \omega + \upsilon \sigma_n^2(|\epsilon_n^{\mathbb{Q}} - \psi_n| - \gamma (\epsilon_n^{\mathbb{Q}} - \psi_n))^2 + \beta \sigma_n^{2}. \nonumber
\end{align}
The computation of the risk-neutral price $C_0^{\mathbb{Q}}$ can be performed with Monte Carlo simulations.

\subsection{Merton jump-diffusion}
For this model, the change of measure used is the one originally proposed in \cite{merton1976option} which assumes no risk premia for jump risk: parameters $\{\mu_J, \sigma_J, \lambda, \sigma\}$ are left unchanged, and the drift parameter $\upsilon$ is shifted to the annualized continuously compounded risk-free rate $r$. The $\mathbb{Q}$-dynamics is 
$$y_n = \left(r - \lambda(e^{\mu_J + \sigma_J^2/2}-1) - \frac{\sigma^{2}}{2}\right)\Delta + \sigma \sqrt{\Delta}\epsilon_n^{\mathbb{Q}} + \sum_{j=N_{n-1}+1}^{N_{n}} \zeta_j, $$
where $\{N_n\}_{n=0}^{N}$ and $\{\zeta_j\}_{j \geq 1}$ have the same distribution than under the physical measure. The risk-neutral price of put options $C_0^{\mathbb{Q}}$ can be computed with the well-known closed-form solution.

\end{document}